%% file: total.tex
\documentclass[%
 reprint,
 amsmath, amssymb,
 aps, prb,
]{revtex4-2}

\usepackage{graphicx}% Include figure files
\usepackage{dcolumn}% Align table columns on the decimal point
\usepackage{bm}% bold math
\usepackage[T1]{fontenc}
\usepackage{amsmath}
\usepackage{amsthm}
\usepackage{amssymb}
\usepackage{amsfonts}
\usepackage{bbm}
\usepackage{dsfont}
\usepackage{relsize}
\usepackage{stmaryrd}
\usepackage{lmodern}
\usepackage{slantsc}
\usepackage{scalefnt}
\usepackage{xspace}
\usepackage{boxedminipage}
\usepackage{ifpdf}
\usepackage{multirow}
\usepackage{hhline}
\usepackage{xifthen}
\usepackage{tensor}
\usepackage{array}
\usepackage{nicematrix}
\usepackage{tabu}
\usepackage{pbox}
\usepackage{tikz}
\usepackage{overpic}
\usepackage{color}
\usepackage{diagbox}
\usepackage{makecell}
\usepackage[mathcal]{euscript}
\usepackage{booktabs}
\usepackage{autobreak}
\usepackage{subfigure}
\usepackage{mathrsfs}
\usetikzlibrary{arrows, matrix, calc, scopes, decorations.markings}

\newtheorem*{ansatz*}{Ansatz}

\newcommand{\ket}[1]{\left\vert{#1}\right\rangle}
\newcommand{\Z}{\mathbb{Z}}

\newcommand{\hs}{{\ }}

\newcommand{\vlc}[8]{{\left[\begin{array}{cc|c} {#1} & {#2\ } & {\ #3}\\ {#4} & {#5\ } & {\ #6}\\ \end{array}\right]^{#7}_{#8}}}

\usetikzlibrary{shapes.geometric, snakes}
\allowdisplaybreaks
\tikzset{>=latex}
\usetikzlibrary{arrows, matrix, calc, scopes, decorations.markings}
\usetikzlibrary{decorations.pathmorphing, positioning}
\tikzset{snake it/.style={decorate, decoration={snake,amplitude=0.2mm,segment length=1mm}}}
\tikzset{->-/.style={decoration={markings, mark=at position .5*\pgfdecoratedpathlength+2pt with {\arrow{>}}},postaction={decorate}}}
\tikzset{-<-/.style={decoration={markings, mark=at position .5*\pgfdecoratedpathlength+2pt with {\arrow{<}}},postaction={decorate}}}
\newtabulinestyle{dash=off 2pt}

\newcolumntype{C}{>{\centering\arraybackslash} m{1.5em} }
\makeatletter
\newcommand*{\Relbarfill@}{\arrowfill@\Relbar\Relbar\Relbar}
\newcommand*{\xeq}[2][]{\ext@arrow 0055\Relbarfill@{#1}{#2}}
\makeatother
\allowdisplaybreaks[4]
\input{totalinput.tex}

\begin{document}

\preprint{APS/123-QED}

\title{Symmetry Fractionalized (Irrationalized) Fusion Rules\\ and Two Domain-Wall Verlinde Formulae}% Force line breaks with \\

\author{Yu Zhao}
%\email{yuzhao20@fudan.edu.cn}
\author{Hongyu Wang}%
%\email{wanghy17@fudan.edu.cn}
\author{Yidun Wan}
\email{ydwan@fudan.edu.cn}
\affiliation{State Key Laboratory of Surface Physics, Department of Physics, Center for Field Theory and Particle Physics, and Institute for Nanoelectronic Devices and Quantum Computing, Fudan University, Shanghai 200433, China.}
\affiliation{Shanghai Qi Zhi Institute, Shanghai 200030, China}
\author{Yuting Hu}
\email{yuting.phys@gmail.com}
\affiliation{School of Physics, Hangzhou Normal University, Hangzhou 311121, China}

\date{\today}

\begin{abstract}
We investigate the composite systems consisting of topological orders separated by gapped domain walls. We derive a pair of domain-wall Verlinde formulae, that elucidate the connection between the braiding of interdomain excitations labeled by pairs of anyons in different domains and quasiparticles in the gapped domain wall with their respective fusion rules. Through explicit non-Abelian examples, we showcase the calculation of such braiding and fusion, revealing that the fusion rules for interdomain excitations are generally fractional or irrational. By investigating the correspondence between composite systems and anyon condensation, we unveil the reason for designating these fusion rules as symmetry fractionalized (irrationalized) fusion rules. Our findings hold promise for applications across various fields, such as topological quantum computation, topological field theory, and conformal field theory.
\end{abstract}


\maketitle

%\tableofcontents

\flushbottom

\section{Introduction}

Topological orders have been extensively studied and significantly influenced our understanding of quantum phases of matter, due to their exotic and intricate properties, such as fusion and braiding of anyons. Anyon fusion is an ultralocal phenomenon that can hardly be directly detected, while braiding is nonlocal and directly measurable\cite{Wen1990a, keski1993, zhang2013, Cincio2013, Lan2014b}. The fusion rule and braiding of anyons are related through the Verlinde formula\cite{verlinde1988}. This formula is the very Verlinde formula in conformal field theory due to the correspondence between topological orders and conformal field theory\cite{pradisi1996, petkova2001, petkova2002, gaberdiel2002, fuchs2008, gaiotto2014a, chang2019}.

While single topological orders in 2+1D systems have been rigorously explored, composite topological systems consisting of multiple topological orders separated by gapped domain walls remain largely uncharted. Prior research\cite{zhao2022} has brought forward the characteristic properties of such composite systems, notably, the nontrivial braiding between quasiparticles in the gapped domain wall and interdomain excitations labeled by pairs of anyons in different domains (see Fig. \ref{fig:domain-wall}). It is crucial to understand how this domain-wall $S$ matrix provides key insights into the fundamental structure of composite topological systems, as well as the relationships among various topological orders.

Moreover, topological orders have been effectively used in quantum computation due to their robustness against local perturbations and decoherence \cite{Iqbal2023, Google2023}. Recent literature has suggested that topological defects could be a more viable candidate for implementing universal computing\cite{Cong2016a, Cong2017b, Cong2017c, Luo2018, Fan2022}. As a generalization of topological boundaries, gapped domain walls have richer properties and thus would also apply to topological quantum computation. Therefore, exploring the properties of these novel excitations in composite systems mentioned above holds immense practical significance.

In this paper, we discover two \emph{domain-wall Verlinde formulae} \eqref{eq:verlinde:id} and \eqref{eq:verlinde:dw} for the composite systems consisting of two topological orders. These two formulae establish the fusion rules of the interdomain excitations as well as the fusion rules of the domain-wall quasiparticles via the domain-wall $S$ matrix, directly generalizing the defect Verlinde formula that relates the fusion and half-linking of boundary excitations in Ref. \cite{Shen2019}. 

In contrast to the conventional fusion rules, where fusion coefficients are always natural numbers, we discover that interdomain excitations can exhibit fractional fusion rules. These fractionalized fusion rules arise because when anyons carrying internal gauge charges in one domain cross the gapped domain wall into another domain, they may become distinct anyons therein, as these indistinguishable internal gauge charges become global symmetry charges. Consequently, the originally unobservable fractional fusion rules of internal gauge charges of certain anyons become the fractionalized fusion rules of interdomain excitations, which are now physical observables. This phenomenon bears a spatial analogy to symmetry breaking in phase transitions triggered by anyon condensation, so we refer to these unique interdomain fusion rules as \emph{symmetry fractionalized fusion rules}.

More surprisingly, the interdomain fusion rules can even be irrational. We believe that these irrationalized fusion rules reflect the algebraic symmetries beyond group description\cite{Hung2013, Gaiotto2014, Gaiotto2019, Gaiotto2020, levin2020, ji2020, kong2020d, kong2022, chatterjee2022}. As the full nature of algebraic symmetries remains enigmatic, our insights into irrationalized fusion rules  provide a clear lens for examining the characteristics of these emergent symmetries.

\section{Domain-wall $S$-matrix $S^\text{DW}$\label{sec:braiding}}

\begin{figure}
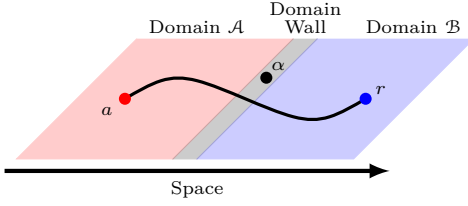
\centering
\FigureDomainWall
\caption{\label{fig:domain-wall}A composite system of two topological orders $\mathcal{A}$ (red) and $\mathcal{B}$ (blue) separated by a gapped domain wall (gray), in which are a domain-wall quasiparticle $\alpha$ and an interdomain excitation with anyon $a$ ($r$) in phase $\mathcal{A}$ ($\mathcal{B}$).}
\end{figure}

We consider a composite system consisting of two different topological orders (domains), denoted by $\mathcal{A}$ and $\mathcal{B}$, separated by a gapped domain wall (See Fig. \ref{fig:domain-wall})\cite{Bais2009, Kitaev2012, Gaiotto2012, HungWan2014, Lan2014, Bais2015, HungWan2015a, Wan2017, Bao2018, Lan2020, jia2022}. In such a system, there can be \emph{domain-wall quasiparticles} and \emph{interdomain elementary excitations}, each bearing a pair $(a,r)$ of anyons $a$ and $r$ respectively in domains $\mathcal{A}$ and $\mathcal{B}$ \cite{zhao2022}, as illustrated in Fig. \ref{fig:domain-wall}.

The gapped domain wall causes a selection rule of allowed interdomain elementary excitations $(a, r)$, recorded in the \emph{branching matrix} $B^{\mathcal{B}\mathcal{A}}$, whose element  $B^{\mathcal{B}\mathcal{A}}_{ra} = 1$ if and only if anyon $a$ in domain $\mathcal{A}$ can enter and become anyon $r$ in $\mathcal{B}$; otherwise, $B^{\mathcal{B}\mathcal{A}}_{ra} = 0$\cite{CFTbook, Lan2014, HungWan2015a}. We find that $B^{\mathcal{B}\mathcal{A}}$ can factorize into two matrices $B^{\mathcal{A}}$ and $B^{\mathcal{B}}$: $B^{\mathcal{A}}_{\alpha a} = 1$ ($B^{\mathcal{B}}_{r\alpha} = 1$) if and only if $\mathcal{A}$-anyon $a$ (domain-wall quasiparticle $\alpha$) can become domain-wall quasiparticle $\alpha$ ($\mathcal{B}$-anyon $r$) when entering the gapped domain wall (domain $\mathcal{B}$) and otherwise $0$. Hence, an interdomain elementary excitation species $(a, r)$ exists if and only if $B^{\mathcal{B}\mathcal{A}}_{ra} = \sum_{\alpha\in\mathcal{L}_\text{DW}}B^{\mathcal{B}}_{r\alpha}B^{\mathcal{A}}_{\alpha a}\ne 0$, where $\mathcal{L}_\text{DW}$ denotes the set of domain-wall quasiparticle species. There are as many domain-wall quasiparticle species as interdomain excitation species\cite{zhao2022}.
% ! A float is stuck (cannot be placed); try class option [floatfix].

\begin{figure}\centering
\subfigure[]{\raisebox{29pt}{\FigureBraiding}\label{fig:braiding:a}}\hspace{5pt}
\subfigure[]{\begin{overpic}[width=3.5cm]{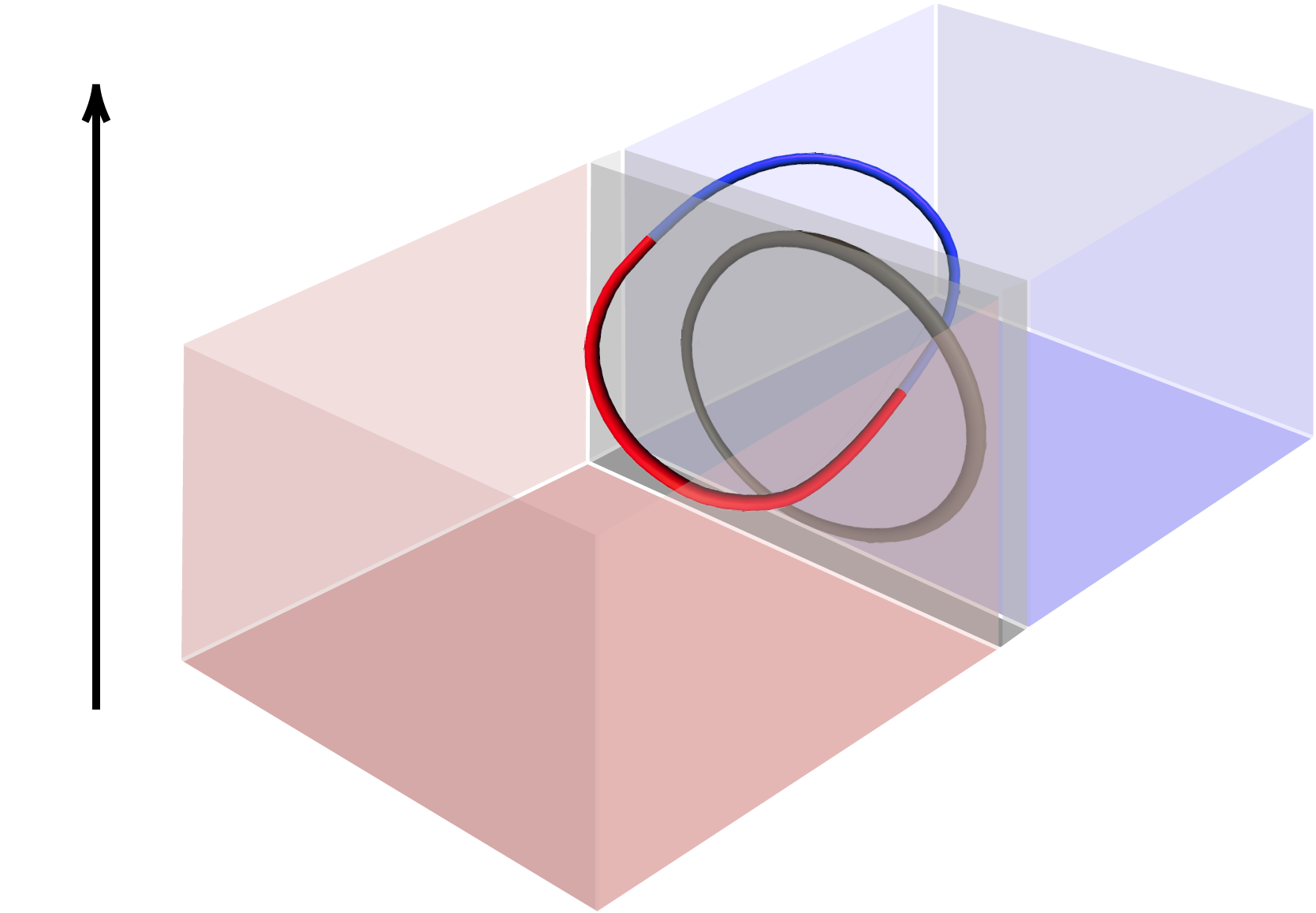}
    \put(39,35){\scriptsize $a$}
    \put(70,58){\scriptsize $r$}
    \put(72,26){\scriptsize $\alpha$}
    \put(10,54){\scriptsize {Time}}
\end{overpic}\label{fig:braiding:b}}
\caption{(a) Braiding interdomain excitation $(a, r)$ around domain-wall quasiparticle $\alpha$, recorded in the $S$-matrix $S^\text{DW}$ up to normalization. Here $1$ labels the trivial domain-wall quasiparticle, and $(1, 1)$ labels the trivial interdomain excitation. (b) Representation of (a) in spacetime.}
\label{fig:braiding}
\end{figure}
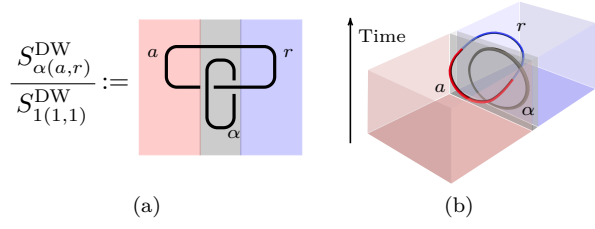

An interdomain excitation may braid nontrivially with a domain-wall quasiparticle. The braiding is encoded in an invertible domain-wall $S$-matrix\cite{zhao2022} $S^\text{DW}$ defined in Fig. \ref{fig:braiding:a}, which is understood (Fig. \ref{fig:braiding:b}) as the linking of the (spacetime) Wilson loops of interdomain excitation $(a, r)$ and domain-wall quasiparticle $\alpha$.

Domain-wall $S$-matrices $S^\text{DW}$ have been computed in the lattice model of such composite systems for doubled topological orders\cite{zhao2022}. Here, we can compute $S^\text{DW}$ from the branching matrices $B^{\mathcal{A}}$ and $B^{\mathcal{B}}$ and the $S$-matrices $S^\mathcal{A}$ and $S^\mathcal{B}$ of $\mathcal{A}$ and $\mathcal{B}$, which may not be doubled, because of the following commutativities we found.
\begin{equation}
\begin{aligned}
\sum_{(a,r)\in\mathcal{L}_\text{ID}}S^\text{DW}_{\alpha(a,r)}\delta_{ab} &= \sum_{c\in\mathcal{L}_{\mathcal{A}}}B^{\mathcal{A}}_{\alpha c}S^{\mathcal{A}}_{cb},\\
\sum_{\alpha\in\mathcal{L}_\text{DW}}B^{\mathcal{B}}_{s\alpha}S^\text{DW}_{\alpha(a,r)} &= \sum_{t\in\mathcal{L}_{\mathcal{B}}}S^{\mathcal{B}}_{st}\delta_{tr},
\end{aligned}
\label{eq:Smatrix:calc}
\end{equation}
where $\mathcal{L}_{\mathrm{ID}}$ and $\mathcal{L}_{\mathcal{A}}$ ($\mathcal{L}_{\mathcal{B}}$) denote the set of interdomain excitation species and the set of anyon species of phase $\mathcal{A}$ ($\mathcal{B}$). Commutativities \eqref{eq:Smatrix:calc} lead to the commutativity found in Ref.\cite{Lan2014, HungWan2015a, CFTbook, Fuchs1992}:
\begin{equation*}
B^{\mathcal{B}\mathcal{A}}S^{\mathcal{A}} = S^{\mathcal{B}}B^{\mathcal{B}\mathcal{A}}.
\end{equation*}

\begin{figure}
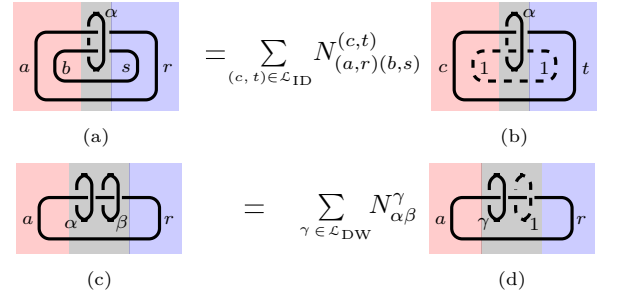
\centering
\subfigure{\FigureVerlindeA\label{fig:verlinde:a}}
\subfigure{\FigureVerlindeB\label{fig:verlinde:b}}\\ \vspace{-1em}
\subfigure{\FigureVerlindeC\label{fig:verlinde:c}}
\subfigure{\FigureVerlindeD\label{fig:verlinde:d}}
\caption{(a) $S^\text{DW}_{\alpha(a,r)}S^\text{DW}_{\alpha(b,s)}$, braiding $(a,r)$ and $(b, s)$ around $\alpha$. (b) $S^\text{DW}_{\alpha(c,t)}S^\text{DW}_{\alpha(1,1)}$, resulted from (a) by fusing $(a, r)$ and $(b, s)$ to $(c, t)$. (c) $S^\text{DW}_{\alpha(a,r)}S^\text{DW}_{\beta(a,r)}$, braiding $(a, r)$ around $\alpha$ and $\beta$. (d) $S^\text{DW}_{\gamma(a,r)}S^\text{DW}_{1(a,r)}$, resulted from (c) by fusing $\alpha$ and $\beta$ to $\gamma$.} 
\label{fig:verlinde}
\end{figure}

Consider for example the doubled Ising and $\Z_2$ toric code phases sandwiching a gapped domain wall. The branching matrices are
% ! Command \tiny invalid in math mode.
\begin{align}
    B^\text{DI} = \tiny{\begin{pmatrix}
    1 & 1 & 0 & 0 & 0 & 0 & 0 & 0 & 0\\
    0 & 0 & 1 & 1 & 0 & 0 & 0 & 0 & 0\\
    0 & 0 & 0 & 0 & 1 & 0 & 0 & 0 & 0\\
    0 & 0 & 0 & 0 & 1 & 0 & 0 & 0 & 0\\
    0 & 0 & 0 & 0 & 0 & 1 & 1 & 0 & 0\\
    0 & 0 & 0 & 0 & 0 & 0 & 0 & 1 & 1
    \end{pmatrix}},
    B^\text{TC} = \tiny{\begin{pmatrix}
    1 & 0 & 0 & 0 & 0 & 0\\
    0 & 1 & 0 & 0 & 0 & 0\\
    0 & 0 & 1 & 0 & 0 & 0\\
    0 & 0 & 0 & 1 & 0 & 0
    \end{pmatrix}}.\nonumber\\
\label{eq:branching}
\end{align}
The branching matrices \eqref{eq:branching} tell that the system has six species of domain-wall quasiparticles, denoted by
\begin{equation*}
1,\quad\quad \epsilon,\quad\quad m,\quad\quad e,\quad\quad \chi,\quad\quad\bar\chi,
\end{equation*}
and six species of interdomain excitations, denoted by
\begin{equation*}
(1\bar 1, 1),\ \ (\psi\bar\psi, 1),\ \ (\psi\bar 1,\epsilon),\ \ (1\bar\psi, \epsilon),\ \ (\sigma\bar\sigma, m),\ \ (\sigma\bar\sigma, e).
\end{equation*}
The matrix $S^\text{DW}$ reads
% ! Command \footnotesize invalid in math mode.
\begin{equation*}
S^\text{DW} = \footnotesize{\frac{1}{2}\begin{pmatrix}
1 & 1 & 1 & 1 & 1 & 1\\
1 & 1 & 1 & 1 & -1 & -1\\
1 & 1 & -1 & -1 & 1 & -1\\
1 & 1 & -1 & -1 & -1 & 1\\
\sqrt{2} & -\sqrt{2} & -\sqrt{2} & \sqrt{2} & 0 & 0\\
\sqrt{2} & -\sqrt{2} & \sqrt{2} & -\sqrt{2} & 0 & 0\\
\end{pmatrix}},
\end{equation*}
which equals the result obtained by the lattice model\cite{zhao2022}.

Now consider a composite chiral system that lacks a lattice-model description: the $\text{su}(2)_{10}$ and $\text{so}(5)_1$ phases separated by a gapped domain wall. In such a system, there are $6$ domain-wall quasiparticle species, denoted by
\begin{equation*}
1,\quad\quad\psi,\quad\quad\sigma,\quad\quad u,\quad\quad u\psi,\quad\quad u\sigma,
\end{equation*}
and $6$ interdomain excitation species, denoted by
\begin{equation*}
(0, 1),\quad (6, 1),\quad (4, \psi),\quad (10, \psi),\quad (3, \sigma),\quad (7,\sigma).
\end{equation*}
The $S^\text{DW}$ matrix then reads
% ! Command \tiny invalid in math mode.
\begin{align*}
S^\text{DW} =
\tiny{\frac{1}{2}\begin{pmatrix}
1 & 1 & 1 & 1 & \sqrt{2} & \sqrt{2}\\
1 & 1 & 1 & 1 & -\sqrt{2} & -\sqrt{2}\\
\sqrt{2} & \sqrt{2} & -\sqrt{2} & -\sqrt{2} & 0 & 0\\
\frac{\sqrt{2} +\sqrt{6}}{2} & \frac{\sqrt{2} -\sqrt{6}}{2} & \frac{-\sqrt{2} +\sqrt{6}}{2} & -\frac{\sqrt{2} +\sqrt{6}}{2} & \sqrt{2} & -\sqrt{2}\\
\frac{\sqrt{2} +\sqrt{6}}{2} & \frac{\sqrt{2} -\sqrt{6}}{2} & \frac{-\sqrt{2} +\sqrt{6}}{2} & -\frac{\sqrt{2} +\sqrt{6}}{2} & -\sqrt{2} & \sqrt{2}\\
1+\sqrt{3} & 1-\sqrt{3} & 1-\sqrt{3} & 1+\sqrt{3} & 0 & 0\\
\end{pmatrix}}.
\end{align*}

\section{Domain-Wall Verlinde Formulae\label{sec:verlinde}}

In a single topological order, e.g., $\mathcal{A}$, the fusion rule $N^{\mathcal{A}}$ and the $S$-matrix $S^{\mathcal{A}}$ satisfy the Verlinde formula\cite{verlinde1988}:
\begin{equation}
\left(N^{\mathcal{A}}\right)_{ab}^c\hs = \sum_{e\in\mathcal{L}_{\mathcal{A}}}\frac{S^{\mathcal{A}}_{ae}S^{\mathcal{A}}_{be}\left(S^{\mathcal{A}}\right)^{-1}_{ec}}{S^{\mathcal{A}}_{1e}} .
\label{eq:verlinde:phase}
\end{equation}
A natural question is: Is there a similar formula for our generalized $S$-matrix $S^\text{DW}$? Indeed, we can imitate formula \eqref{eq:verlinde:phase} to define fusion rules $N_{(a,r)(b,s)}^{(c,t)}$ for interdomain excitations and $N_{\alpha\beta}^\gamma$ for domain-wall quasiparticles via our domain-wall $S$-matrix $S^{\text{DW}}$:
\begin{equation}
N_{(a,r)(b,s)}^{(c,t)}\hs := \sum_{\alpha\in\mathcal{L}_\text{DW}}\frac{S^\text{DW}_{\alpha(a,r)}S^\text{DW}_{\alpha(b,s)}\left(S^\text{DW}\right)^{-1}_{(c,t)\alpha}}{S^\text{DW}_{\alpha(1,1)}} ,
\label{eq:verlinde:id}
\end{equation}
\begin{equation}
N_{\alpha\beta}^{\gamma}\hs := \sum_{(a,r)\in\mathcal{L}_\text{ID}}\frac{S^\text{DW}_{\alpha(a,r)}S^\text{DW}_{\beta(a,r)}\left(S^\text{DW}\right)^{-1}_{(a,r)\gamma}}{S^\text{DW}_{1(a,r)}} .
\label{eq:verlinde:dw}
\end{equation}

Multiplying formula \eqref{eq:verlinde:id} by $S^\text{DW}_{\alpha(c,t)}$ and then summing over $(c,t)\in \mathcal{L}_\text{ID}$ results in
\begin{equation*}
\frac{S^\text{DW}_{\alpha(a,r)}}{S^\text{DW}_{1(1,1)}}\hs \frac{S^\text{DW}_{\alpha(b,s)}}{S^\text{DW}_{1(1,1)}}\hs = \sum_{(c,t)\in\mathcal{L}_\text{ID}}N_{(a,r)(b,s)}^{(c,t)}\hs \frac{S^\text{DW}_{\alpha(c,t)}}{S^\text{DW}_{1(1,1)}}\hs \frac{S^\text{DW}_{\alpha(1,1)}}{S^\text{DW}_{1(1,1)}}\hs ,
\end{equation*}
which is understood in Fig. \ref{fig:verlinde:a} and \ref{fig:verlinde:b}. Hence, $N_{(a,r)(b,s)}^{(c,t)}$ are justified as the fusion coefficients for interdomain excitations. Similarly, $N_{\alpha\beta}^{\gamma}$ defined in formula \eqref{eq:verlinde:dw} are indeed the domain-wall fusion coefficients, as understood in Fig. \ref{fig:verlinde:c} and \ref{fig:verlinde:d}. It can be shown that $N_{\alpha\beta}^{\gamma}$ represents the number of channels that two domain-wall quasiparticles $\alpha$ and $\beta$ fuse to quasiparticle $\gamma$. Therefore, formulae \eqref{eq:verlinde:id} and \eqref{eq:verlinde:dw} --- now called the \emph{domain-wall Verlinde formulae} --- generalize the Verlinde formula in a single topological order and the defect Verlinde fromula\cite{Shen2019}.

\section{Interdomain Fusion Algebra\\ and Quantum Dimensions\label{sec:algebra}}

\begin{figure}
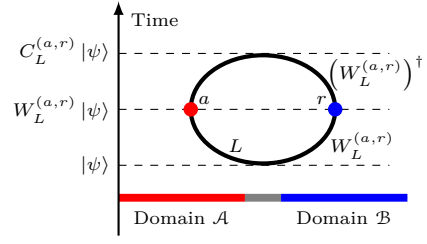

\FigureLoopOperator
\caption{Interdomain string operator $W_L^{(a, r)}$ and interdomain loop operator $C_L^{(a,r)}$ acting on state $\ket\psi$.}
\label{fig:loopoperator}
\end{figure}
    
Interdomain excitations extend the notion of anyons in a single topological order, which have quantum dimensions that form a $1$-dimensional representation of the fusion algebra of the anyons. Do interdomain excitations also have fusion algebra and quantum dimensions? 

In an arbitrary state $\ket{\psi}$ of a composite system, we can define a string operator $W_L^{(a, r)}$ that creates an interdomain excitation $(a, r)$, with anyons $a$ and $r$ at the two ends of an interdomain path $L$ (Fig. \ref{fig:loopoperator}). Along a fixed path $L$, all string operators $W_L^{(a, r)}$ form an algebra:
\begin{equation}
W_L^{(a, r)}W_L^{(b,s)} = \sum_{c}f_{(a,r)(b,s)}^{(c,t)}W_L^{(c,t)}.
\label{eq:string}
\end{equation}
Interdomain fusion coefficients $N_{(a,r)(b,s)}^{(c,t)}$ can be expressed in terms of the operator product expansion coefficients $f_{(a,r)(b,s)}^{(c,t)}$:
\begin{equation}
N_{(a,r)(b,s)}^{(c,t)} = \left|f_{(a,r)(b,s)}^{(c,t)}\right|^2.
\label{eq:algebra}
\end{equation}

To justify this equation, we define the interdomain loop operators (Fig. \ref{fig:loopoperator}):
\begin{equation}
C_{L}^{(a, r)} := \left(W_{L}^{(a, r)}\right)^\dagger W_{L}^{(a, r)}.
\label{eq:loopoperator}
\end{equation}
Compared to the Fig. \ref{fig:verlinde:a} and \ref{fig:verlinde:b}, the interdomain fusion coefficients $N_{(a,r)(b,s)}^{(c,t)}$ are the operator product expansion coefficients of interdomain loop operators:
\begin{equation}
C_{L}^{(a, r)}C_{L}^{(b, s)} = \sum_{(c,t)\in\mathcal{L}_\text{ID}}N_{(a,r)(b,s)}^{(c,t)}C_{L}^{(c, t)},
\label{eq:loop}
\end{equation}
which defines the fusion algebra of interdomain excitations. Equations \eqref{eq:string}, \eqref{eq:loopoperator}, and \eqref{eq:loop} lead to Eq. \eqref{eq:algebra}.

We define the \emph{quantum dimension} $d_{(a,r)}$ of interdomain excitation $(a, r)$ by
\begin{equation}
d_{(a, r)} := \left\langle 0\middle|C_L^{(a, r)}\middle|0\right\rangle = \frac{S^\text{DW}_{1(a,r)}}{S^\text{DW}_{1(1, 1)}}\hs ,
\label{eq:quantumdimension}
\end{equation}
where $\ket 0$ is the vacuum state, and the equality follows that the trivial domain-wall quasiparticle $1$ braids trivially with interdomain excitations. Definition \eqref{eq:quantumdimension} complies with the definition of quantum dimensions in a single topological order. By Eqs. \eqref{eq:loop} and \eqref{eq:quantumdimension},
\begin{equation}
d_{(a, r)}d_{(b,s)} = \sum_{(c, t)\in\mathcal{L}_\text{ID}}N_{(a,r)(b,s)}^{(c,t)}d_{(c,t)},
\label{eq:representation}
\end{equation}
indicating that the quantum dimensions $d_{(a,r)}$ are a $1$-dimensional representation of the fusion algebra \eqref{eq:loop}. It can be shown that $d_{(a,r)}$ is the largest eigenvalue of matrix $\left[N_{(a, r)}\right]^{(c,t)}_{(b,s)} := N_{(a,r)(b,s)}^{(c, t)}$. And in a system of infinitely many interdomain excitations $(a, r)$, $d_{(a,r)}$ is the asymptotic dimension of the Hilbert space of each $(a,r)$.

When fusing two interdomain excitations $(a, r)$ and $(b, s)$ in state $\ket\psi$, the resulting state $W_L^{(a,r)}W_L^{(b,s)}\ket\psi$ is a superposition of orthogonal interdomain excitation states $W_L^{(c,t)}\ket\psi$. The probability of measuring $(c, t)$ is
\begin{equation}
P_{(c,t)} = \frac{N_{(a,r)(b,s)}^{(c,t)}d_{(c,t)}}{d_{(a,r)}d_{(b,s)}}.
\label{eq:prob}
\end{equation}

Such a composite system, including the interdomain excitations and their fusion, appears to be describable by a fusion 2-category, which we shall report elsewhere.

\section{Symmetry Fractionalized / Irrationalized Fusion Rules\label{sec:symmetry}}

The fusion coefficients in a single topological order are always natural numbers, and so are the domain-wall fusion coefficients $N^\gamma_{\alpha\beta}$. Nevertheless, interdomain fusion coefficients $N_{(a,r)(b,s)}^{(c,t)}$ can be fractional. For instance, in the composite system of the doubled Ising and $\Z_2$ toric code phases, domain-wall Verlinde Formula \eqref{eq:verlinde:id} leads to
\begin{equation}
\begin{aligned}
(\sigma\bar\sigma, e)\times (\sigma\bar\sigma, e) &= \frac{1}{2}(1\bar 1, 1) + \frac{1}{2}(\psi\bar\psi, 1),\\
(\sigma\bar\sigma, m)\times (\sigma\bar\sigma, m) &= \frac{1}{2}(1\bar 1, 1) + \frac{1}{2}(\psi\bar\psi, 1),\\
(\sigma\bar\sigma, e)\times (\sigma\bar\sigma, m) &= \frac{1}{2}(\psi\bar 1, \epsilon) + \frac{1}{2}(1\bar\psi, \epsilon).
\end{aligned}
\label{eq:semi}
\end{equation}
The quantum dimensions are $d_{(\sigma\bar\sigma, e)} = d_{(\sigma\bar\sigma_, m)} = d_{(1\bar 1, 1)} = d_{(\psi\bar\psi, 1)} = d_{(\psi\bar 1, \epsilon)} = d_{(1\bar\psi, \epsilon)} = 1$; therefore, fusion rules \eqref{eq:semi} satisfy Eq. \eqref{eq:representation}. We shall call such fusion rules \emph{symmetry fractionalized fusion rules}. The symmetry fractionalization is understood by the correspondence between the spatial composite system and temporal anyon condensation as follows.

The $\Z_2$ toric code topological order can originate from the doubled-Ising phase via a phase transition triggered by $\psi\bar\psi$ condensation in the doubled-Ising phase, where an anyon $\sigma\bar\sigma$ carries internal $\Z_2$ gauge charges\cite{Hu2017, zhao2022}. We denote $\sigma\bar\sigma$ with $\Z_2$-charge $0$ ($1$) as $\sigma\bar\sigma_0$ ($\sigma\bar\sigma_1$), which can transform into each other by gauge transformations and are thus unobservable. The $\psi\bar\psi$ condensation breaks this $\Z_2$ gauge invariance to the global $\Z_2$ symmetry of the $\Z_2$ toric code phase, such that as also dictated by branching matrix Eq. \eqref{eq:branching}, $\sigma\bar\sigma_0$ ($\sigma\bar\sigma_1$) becomes toric-code anyon $m$ ($e$), which are now topological observables. We can find that $\sigma\bar\sigma_0$ and $\sigma\bar\sigma_1$ follow fractional fusion rules:
\begin{align*}
\sigma\bar\sigma_0\times\sigma\bar\sigma_0 = \sigma\bar\sigma_1&\times\sigma\bar\sigma_1 = \frac{1}{2}1\bar 1 + \frac{1}{2}\psi\bar\psi,\\
\sigma\bar\sigma_0\times\sigma\bar\sigma_1 &= \frac{1}{2}\psi\bar 1 + \frac{1}{2}1\bar\psi,
\label{eq:gauge}
\end{align*}
which are likewise unobservable.

Branching matrix \eqref{eq:branching} establishes the following map:
\begin{equation*}
(\sigma\bar\sigma, m) \longleftrightarrow \sigma\bar\sigma_0,\quad\quad  (\sigma\bar\sigma, e) \longleftrightarrow \sigma\bar\sigma_1.
\end{equation*}
As a result, interdomain excitations $(\sigma\bar\sigma,m)$ and $(\sigma\bar\sigma, e)$ fuse in the same manner as $\sigma\bar\sigma_0$ and $\sigma\bar\sigma_1$. Nonetheless, $m$ and $e$ are topological observables of interdomain excitations; therefore, the fusion rules \eqref{eq:semi} are physically measurable and are thus justified to be called symmetry fractionalized fusion rules.

Fractionalized fusion rules occur often in such composite systems, where one domain could arise from the other via anyon condensation, which usually breaks certain gauge invariance. The broken gauge invariance however may not always be describable by a gauge group but rather by certain algebra, such that the resultant global symmetry is also algebraic\cite{Hung2013, Gaiotto2014, Gaiotto2019, Gaiotto2020, levin2020, ji2020, kong2020d, kong2022, chatterjee2022}; hence, more generally, the interdomain fusion rules can be more complicated than being just fractionalized. For example, in the composite system of $\text{su}(2)_{10}$ and $\text{so}(5)_1$ phases, the fusion coefficients $N_{(a,r)(b,s)}^{(c,t)}$ can be even irrational: 
\begin{align*}
&(6, 1)\times(6, 1) = (2-\sqrt{3})(0, 1) + (\sqrt{3} - 1)(6, 1),\\
&(4, \psi)\times(4, \psi) = (2-\sqrt{3})(0, 1) + (\sqrt{3} - 1)(6, 1),\\ 
&(6, 1)\times(4, \psi) = (2 - \sqrt{3})(10, \psi) + (\sqrt{3} - 1)(4, \psi),\\
&(6, 1)\times(3, \sigma) =\ \frac{\sqrt{3} - 1}{2}(3, \sigma) + \frac{3 - \sqrt{3}}{2}(7, \sigma),\\
&(4, \psi)\times(7, \sigma) =\ \frac{\sqrt{3} - 1}{2}(3, \sigma) + \frac{3 - \sqrt{3}}{2}(7, \sigma),\\
&(4, \psi)\times(3, \sigma) =\ \frac{3 - \sqrt{3}}{2}(3, \sigma) + \frac{\sqrt{3} - 1}{2}(7, \sigma),\\ 
&(6, 1)\times(7, \sigma) =\ \frac{3 - \sqrt{3}}{2}(3, \sigma) + \frac{\sqrt{3} - 1}{2}(7, \sigma),\\
&(3,\sigma)\times (3, \sigma) =\ \left(1 - \frac{\sqrt{3}}{3}\right)(0, 1) + (4,\psi) + \frac{\sqrt{3}}{3}(6, 1),\\
&(7,\sigma)\times(7,\sigma) =\ \left(1 - \frac{\sqrt{3}}{3}\right)(0, 1) + (4,\psi) + \frac{\sqrt{3}}{3}(6, 1),\\
&(3,\sigma)\times (7, \sigma) = \frac{\sqrt{3}}{3}(4, \psi) + (6,1) + \left(1 - \frac{\sqrt{3}}{3}\right)(10, \psi).
\end{align*}
The symmetry irrationalized fusion rules may offer new insight into studying the algebraic global symmetries of topological orders.

\begin{acknowledgments}
YW is supported by the General Program of Science and Technology of Shanghai No. 21ZR1406700, and Shanghai Municipal Science and Technology Major Project (Grant No.2019SHZDZX01). YW is grateful for the Hospitality of the Perimeter Insitute during his visit, where the main part of this work is done. YH is supported by Zhejiang Provincial Natural Science Foundation of China (No. LY23A050001). The authors are grateful to Jiaqi Guo for helping draw Fig. \ref{fig:braiding:b}. The authors also appreciate Davide Gaiotto, Yingcheng Li, Jiaqi Guo, and Yanyan Chen for their helpful discussions. 
\end{acknowledgments}

\appendix

\begin{widetext}

    \section{A Brief Review of $2+1$D Topological Orders\label{appendix:topologicalorder}}
    
    Since our composite system consists of $2+1$D topological orders, we briefly review the related properties of topological order, denoted by $\mathcal{A}$.
    \begin{enumerate}
    \item The fundamental physical data consists of a finite set of distinct anyonic excitation species, denoted by $\mathcal{L}_{\mathcal{A}}$.
    \item In each elementary excitation state $\ket{a}_L$, there are two anyons $a, a^*\in\mathcal L_{\mathcal{A}}$ excited at the ends of a path $L$ in the system by a string operator $W_L^a$ (Fig. \ref{fig:single:a}):
    \begin{equation*}
    \ket a_L = W_L^a\ket 0,
    \end{equation*}
    where $a^*$ is the anti-particle of $a$ and $\ket 0$ is the vacuum state. The path $L$ can be homotopically deformed with its endpoints fixed.
    \item Anyons in $\mathcal{A}$ can fuse. The fusion rule is described by a three-index fusion tensor $N^\mathcal{A}$. The component $\left(N^\mathcal{A}\right)_{ab}^c\in\mathbb{N}$ represents the number of channels where two anyons $a$ and $b$ fuse to anyon $c$, i,e, the dimension of the Hilbert space spanned by states containing three anyons $a^*$, $b^*$, and $c$ (Fig. \ref{fig:single:b}).
    
    Anyon fusion is associative and commutative and can be simply expressed as
    \begin{equation*}
    a\times b = \sum_{c\in\mathcal{L}_\mathcal{A}}N_{ab}^cc.
    \end{equation*}
    
    Fusion coefficients $(N^\mathcal{A})_{ab}^c$ are also the operator product expansion coefficients of loop operators $C_L^a$,
    \begin{equation*}
    C_L^aC_L^b = \sum_{c\in\mathcal{L}_{\mathcal{A}}}N_{ab}^cC_L^c,
    \end{equation*}
    defining the fusion algebra of topological order $\mathcal{A}$. Here,
    \begin{equation*}
    C_L^a := \left(W_L^a\right)^\dagger W_L^a.
    \end{equation*}
    \item The quantum dimension of anyon $a\in\mathcal{L}_{\mathcal{A}}$ is
    \begin{equation*}
    d_a := \left\langle 0\middle|C_L^a\middle|0\right\rangle.
    \end{equation*}
    It can be shown that $d_a$ is the largest eigenvalue of matrix $\left[N_a^{\mathcal{A}}\right]_b^c := \left(N^{\mathcal{A}}\right)_{ab}^c$. Thus, in a system with infinitely many anyons $a$, $d_a$ is the asymptotic dimension of the Hilbert space for each anyon $a$.
    
    Quantum dimensions are a 1-dimensional representation of the fusion algebra:
    \begin{equation*}
    d_ad_b = \sum_{c\in\mathcal{L}_{\mathcal{A}}}N_{ab}^cd_c.
    \end{equation*}
    \item Anyons in $\mathcal{A}$ can braid with each other. The braiding of two anyons $a, b\in\mathcal{L}_{\mathcal{A}}$ is encoded in the modular $S$-matrix $S^{\mathcal{A}}$:
    \begin{equation*}
     \frac{S^\mathcal{A}_{ab}}{S^\mathcal{A}_{11}} := \FigureSingleC\hs,
    \end{equation*}
    where $1\in \mathcal{L}_{\mathcal{A}}$ denotes the trivial anyon. Since the trivial anyon $1$ braids trivially with any anyon,
    \begin{equation*}
    d_a = \frac{S^\mathcal{A}_{1a}}{S^\mathcal{A}_{11}} .
    \end{equation*}
    
    The fusion coefficients and the $S$-matrix satisfy the following \emph{Verlinde Formula}\cite{verlinde1988}:
    \begin{equation*}
    \left(N^\mathcal{A}\right)_{ab}^c = \sum_{e\in\mathcal L_{\mathcal{A}}}\frac{S_{ae}^{\mathcal{A}}S_{be}^{\mathcal{A}}\left[S^{\mathcal{A}}\right]^{-1}_{ec}}{S_{1e}^{\mathcal{A}}}.
    \end{equation*}
    \end{enumerate}
    
    \begin{figure}
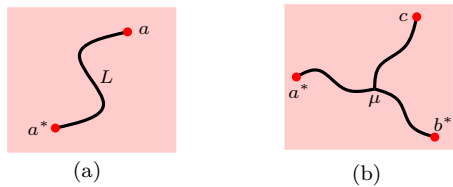
\centering
    \subfigure[]{\FigureSingleA\label{fig:single:a}}\hspace{30pt}
    \subfigure[]{\FigureSingleB\label{fig:single:b}}
    \caption{(a) Elementary excitation state $\ket{a}_L$. At the two ends of path $L$ are two anyons $a^*$ and $a$. (b) Fusing anyons $a$ and $b$ results in a new anyon $c$, where $1\le \mu\le N_{ab}^c$ labels different possible fusion channels.}
    \label{fig:singlephase}
    \end{figure}
    
    \section{A Brief Review of Anyon Condensation\label{appendix:anyoncondensation}}
    
    We briefly review the phase transition from one topological phase, denoted by parent phase $\mathcal{A}$, to another child phase $\mathcal{B}$ triggered by certain anyon condensation in $\mathcal{A}$. An auxiliary intermediate phase $\mathcal{T}$ is introduced merely as a method to study the anyon condensation process\cite{Bais2009a}. The anyon condensation in $\mathcal{A}$ first leads to intermediate phase  $\mathcal{T}$, followed by the transition from the intermediate phase to child phase $\mathcal{B}$ (Fig. \ref{fig:correspondence:a}).
    
    The most important data characterizing anyon condensation is the set of \emph{condensed anyons} in $\mathcal{A}$. These anyons behave like trivial quasiparticles in $\mathcal{T}$, while the other anyons behave like nontrivial quasiparticles. It appears that (i) certain anyons in $\mathcal{A}$, including the condensed ones, may correspond to distinct quasiparticles in $\mathcal{T}$. This phenomenon is called \emph{splitting}. (ii) Two types of $\mathcal{A}$-anyons may correspond to identical $\mathcal{T}$-quasiparticles if they are related by fusing with a condensed anyon in the parent phase. This phenomenon is called \emph{identification}. We find that the relations between the anyons in $\mathcal{A}$ and the quasiparticles in $\mathcal{T}$ can be encoded in the branching matrix $B^\mathcal{A}$: An anyon $a\in\mathcal{L}_{\mathcal{A}}$ behaves like quasiparticle $\alpha$ in the intermediate phase if and only if $B^\mathcal{A}_{\alpha a} = 1$; otherwise, $B^\mathcal{A}_{\alpha a} = 0$.
    
    The quasiparticles in intermediate phase $\mathcal{T}$ can also fuse. The fusion rule is encoded in the fusion tensor $N^{\mathcal{T}}$, whose component $\left(N^{\mathcal{T}}\right)^{\gamma}_{\alpha\beta}$ represents the number of channels that two quasiparticles $\alpha$ and $\beta$ in $\mathcal{T}$ fuse to quasiparticle $\gamma$. The fusion tensor commutes with branching matrix $B^\mathcal{A}$:
    \begin{equation*}
    \sum_{c\in\mathcal{L}_{\mathcal{A}}}B^\mathcal{A}_{\gamma c}\left(N^{\mathcal{A}}\right)^c_{ab} = \sum_{\alpha\beta\in\mathcal{L}_{\mathcal{T}}}\left(N^{\mathcal{T}}\right)_{\alpha\beta}^{\gamma}B^\mathcal{A}_{\alpha a}B^\mathcal{A}_{\beta b},
    \end{equation*}
    where $\mathcal{L}_{\mathcal{T}}$ denotes the set of quasiparticle species in the gapped domain wall.
    
    Not all quasiparticles in $\mathcal{T}$ appear in child phase $\mathcal{B}$. If quasiparticle $\alpha$ originates from an $\mathcal{A}$-anyon $a$ with nontrivial braiding with a condensed anyon $\gamma$, this quasiparticle is said to be \emph{confined} in the child phase. Only unconfined quasiparticles are allowed in the child phase as $\mathcal{B}$-anyons. We define the branching matrix $B^\mathcal{B}$ to record the relations between the quasiparticles in $\mathcal{T}$ and the anyons in $\mathcal{B}$, where $B^\mathcal{B}_{r\alpha} = 1$ if and only if quasiparticle $\alpha$ in the intermediate phase may become anyon $r$ in child phase $\mathcal{B}$.
    
    Composing two branch matrices $B^\mathcal{A}$ and $B^\mathcal{B}$ results in the total branching matrix $B^{\mathcal{B}\mathcal{A}}$ relating the anyon species in $\mathcal{A}$ and $\mathcal{B}$\cite{Fuchs2002, Lan2014, HungWan2015a}:
    \begin{equation*}
    B^{\mathcal{B}\mathcal{A}}_{ra} = \sum_{\alpha\in\mathcal{L}_{\mathcal{T}}}B^\mathcal{B}_{r\alpha}B^\mathcal{A}_{\alpha a}.
    \end{equation*}
    Branching matrix $B^{\mathcal{B}\mathcal{A}}$ commutes with the fusion rules and $S$-matrices:
    \begin{equation*}
    \sum_{c\in\mathcal{L}_{\mathcal{A}}}B^{\mathcal{B}\mathcal{A}}_{tc}\left(N^\mathcal{A}\right)_{ab}^c = \sum_{rs\in\mathcal{L}_{\mathcal{B}}}\left(N^\mathcal{B}\right)_{rs}^t B^{\mathcal{B}\mathcal{A}}_{ra}B^{\mathcal{B}\mathcal{A}}_{sb},\quad\quad\quad\quad\quad\quad BS^\mathcal{A} = S^\mathcal{B}B.
    \end{equation*}
    
    To study anyon condensation, the Hilbert space of child phase $\mathcal{B}$ can be embedded in the Hilbert space of the parent phase $\mathcal{A}$\cite{Eliens2013, Gu2014a}. Specifically, a child excitation state with three anyons $r$, $s$, and $t$ is embedded in the Hilbert space of the parent phase, as a linear combination of excitation states with three parent anyons $a$, $b$, and $c$ (Fig. \ref{fig:correspondence:b}). The composition coefficients, denoted by 
    \begin{equation*}
    \vlc{r}{s}{t}{a}{b}{c}{\mu}{\nu}\hs ,
    \end{equation*}
    are called \emph{vertex lifting coefficients} (VLCs), where $\mu$ and $\nu$ label bases of the parent and child excitation states.
    
    \begin{figure}
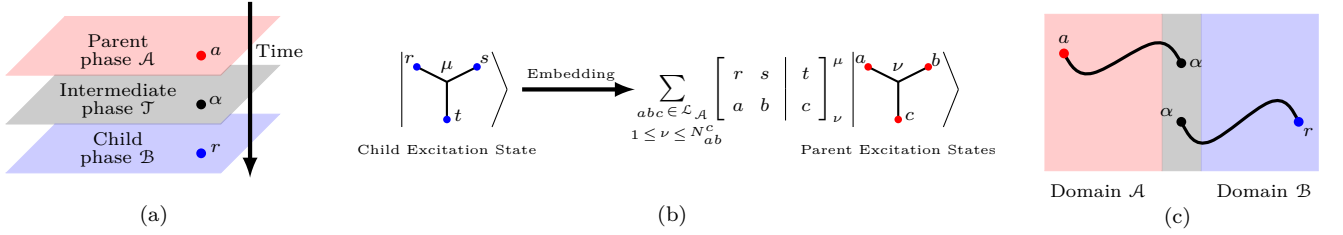
\centering
    \subfigure[]{\FigureCorrespondenceA\label{fig:correspondence:a}}\hspace{5pt}
    \subfigure[]{\FigureCorrespondenceB\label{fig:correspondence:b}}\hspace{5pt}
    \subfigure[]{\FigureCorrespondenceC\label{fig:correspondence:c}}
    \caption{(a) In anyon condensation, anyon $a$ in parent phase $\mathcal{A}$ may be tranformed to quasiparticle $\alpha$ in intermediate phase $\mathcal{T}$, and then become anyon $r$ in child phase $\mathcal{B}$. (b) The embedding from the child Hilbert space to the parent Hilbert space. (c) Elementary excitations $(a, \alpha)$ and $(\alpha, r)$ in the system with topological orders $\mathcal{A}$ and $\mathcal{B}$ separated by a gapped domain wall.}
    \label{fig:correspondence}
    \end{figure}
        
    \section{Correspondence between Spatial Composite Systems and Temporal Anyon Condensation\label{appendix:correspondence}}
    
    \begin{figure}
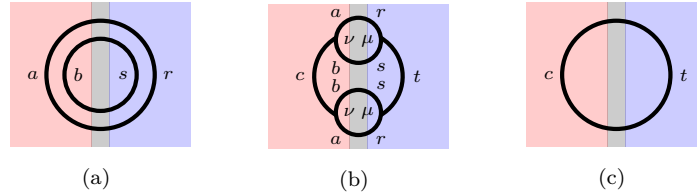
\centering
    \subfigure[]{\FigureVLCA}\hspace{20pt}
    \subfigure[]{\FigureVLCB}\hspace{20pt}
    \subfigure[]{\FigureVLCC}
    \caption{(a) Composing two interdomain loop operators $C_L^{(a,r)}$ and $C_L^{(b,s)}$, results in the linear composition of (c) interdomain loop operators $C_L^{(c,t)}$. (b) is an intermediate state, where each vertex in domain $\mathcal{B}$ can be expressed as a linear combination of vertices in the parent phase $\mathcal{A}$, with vertex lifting coefficients.}
    \label{fig:vlc}
    \end{figure}
    
    The phase transition from parent phase $\mathcal{A}$ to child phase $\mathcal{B}$ is temporal and corresponds to a spatial composite system consisting of $\mathcal{A}$ and $\mathcal{B}$ separated by a gapped domain wall\cite{zhao2022}, as illustrated in Fig. \ref{fig:correspondence}.
    
    The gapped domain wall serves as the spatial counterpart of the auxiliary, virtual intermediate phase during anyon condensation. Each quasiparticle species in the gapped domain wall corresponds one-to-one with the quasiparticle species in the intermediate phase $\mathcal{T}$ during the phase transition. The domain-wall quasiparticles and the quasiparticles in intermediate phase $\mathcal{T}$ share the same fusion rules.
    
    An anyon $a$ in domain $\mathcal{A}$ can always enter the gapped domain wall and become a domain-wall quasiparticle, say $\alpha$; but $\alpha$ may not be able to enter domain $\mathcal{B}$. If $\alpha$ can enter $\mathcal{B}$, it would become anyon $r$ in $\mathcal{B}$. This process, where an $\mathcal{A}$-anyon $a$ becomes quasiparticle $\alpha$ and then $r$ when moving from domain $\mathcal{A}$ across the gapped domain wall into domain $\mathcal{B}$, corresponds to the transformation in anyon condensation, where $\mathcal{A}$-anyon $a$ first becomes quasiparticle $\alpha$ in the intermediate phase and then $r$ in the child phase\cite{zhao2022}. Hence, the gapped domain wall imposes a selection rule for anyon transformations when entering different domains, which is encoded in the same branching matrices $B^\mathcal{A}$, $B^\mathcal{B}$, and $B^\mathcal{BA}$, as in anyon condensation.
    
    We find that in a composite system corresponding to an anyon-condensation-induced phase transition, $d_{(a, r)} = d_r$, and the interdomain fusion rules can be written in terms of vertex lifting coefficients of anyon condensation (Fig. \ref{fig:vlc}):
    \begin{equation}
    N_{(a, r) (b,s)}^{(c, t)} = \sqrt{\frac{d_rd_sd_c}{d_ad_bd_t}}\sum_{\mu\nu}\left|\vlc{r}{s}{t}{a}{b}{c}{\mu}{\nu}\right|^2.
    \end{equation}
    
    \section{Examples\label{appendix:examples}}
    
    We apply our methods in the article to computing the domain-wall $S$-matrices, interdomain fusion rules, and domain-wall fusion rules of specific composite systems of two topological orders separated by a gapped domain wall.
    
    \subsection{The composite system of the doubled-Ising and $\Z_2$ toric code phases\label{appendix:ditc}}
    
    Consider the composite system consisting of the doubled-Ising and $\Z_2$ toric code phases separated by a gapped domain wall. There are $9$ anyon species in the doubled-Ising phase:
    \begin{equation*}
        1\overline{1},\quad\quad 1\overline{\sigma},\quad\quad 1\overline{\psi},\quad\quad \sigma\overline{1},\quad\quad \sigma\overline{\sigma},\quad\quad \sigma\overline{\psi},\quad\quad \psi\overline{1},\quad\quad \psi\overline{\sigma},\quad\quad \psi\bar\psi,
    \end{equation*}
    whose quantum dimensions are $d_{1\bar 1} = d_{\psi\bar\psi} = d_{\psi\bar 1} = d_{1\bar\psi} = 1$, $d_{\sigma\bar 1} = d_{1\bar\sigma} = d_{\sigma\bar\psi} = d_{\psi\bar\sigma} = \sqrt{2}$, and $d_{\sigma\bar\sigma} = 2$. The $S$-matrix of the doubled-Ising phase is 
    \begin{equation*}
    S^\text{DI} \hs = \hs \frac{1}{4}\quad \begin{pNiceMatrix}[first-row,first-col]
        & 1\bar{1} & \psi\bar\psi & \psi\bar{1} & 1\bar{\psi} & \sigma\bar{\sigma} & \sigma\bar 1 & \sigma\bar\psi & 1\bar \sigma & \psi\bar\sigma\\
        1\bar 1 & 1 & 1 & 1 & 1 & 2 & \sqrt{2} & \sqrt{2} & \sqrt{2} & \sqrt{2}\\
        \psi\bar\psi & 1 & 1 & 1 & 1 & 2 & -\sqrt{2} & -\sqrt{2} & -\sqrt{2} & -\sqrt{2}\\
        \psi\bar 1 & 1 & 1 & 1 & 1 & -2 & -\sqrt{2} & -\sqrt{2} & \sqrt{2} & \sqrt{2}\\
        1\bar\psi & 1 & 1 & 1 & 1 & -2 & \sqrt{2} & \sqrt{2} & -\sqrt{2} & -\sqrt{2}\\
        \sigma\bar\sigma & 2 & 2 & -2 & -2 & 0 & 0 & 0 & 0 & 0\\
        \sigma\bar 1 & \sqrt{2} & -\sqrt{2} & -\sqrt{2} & \sqrt{2} & 0 & 0 & 0 & 2 & -2\\
        \sigma\bar\psi & \sqrt{2} & -\sqrt{2} & -\sqrt{2} & \sqrt{2} & 0 & 0 & 0 & -2 & 2\\
        1\bar\sigma & \sqrt{2} & -\sqrt{2} & \sqrt{2} & -\sqrt{2} & 0 & 2 & -2 & 0 & 0\\
        \psi\bar\sigma & \sqrt{2} & -\sqrt{2} & \sqrt{2} & -\sqrt{2} & 0 & -2 & 2 & 0 & 0\\
        \end{pNiceMatrix}.
    \end{equation*}
    
    Condensing anyon $\psi\bar\psi$ in the doubled-Ising phase results in the $\Z_2$ toric code phase. The anyon species of $\Z_2$ toric code phase are $1, e, m, \epsilon$, whose quantum dimensions are all $1$. The $S$-matrix of $\Z_2$ toric code phase is
    \begin{equation*}
    S^\text{TC} \hs = \hs \frac{1}{2}\quad \begin{pNiceMatrix}[first-row,first-col]
        & 1 & \epsilon & m & e\\
        1 & 1 & 1 & 1 & 1 \\
        \epsilon & 1 & 1 & -1 & -1 \\
        m & 1 & -1 & 1 & -1 \\
        e & 1 & -1 & -1 & 1 \\
        \end{pNiceMatrix}.
    \end{equation*}
    
    In the composite system of the doubled-Ising and $\Z_2$ toric code phases separated by a gapped domain wall, there are $6$ domain-wall quasiparticle species:
    \begin{equation*}
    1,\quad\quad e,\quad\quad m,\quad\quad \epsilon,\quad\quad \chi, \quad\quad \bar\chi.
    \end{equation*}
    The quantum dimensions are $d_1 = d_\epsilon = d_e = d_m = 1$, and $d_\chi = d_{\bar\chi} = \sqrt{2}$. The branch matrices are
    \begin{equation*}
    B^\text{DI} \hs =\hs \begin{pNiceMatrix}[first-row,first-col]
        & 1\bar{1} & \psi\bar\psi & \psi\bar{1} & 1\bar{\psi} & \sigma\bar{\sigma} & \sigma\bar 1 & \sigma\bar\psi & 1\bar \sigma & \psi\bar\sigma\\
        1 & 1 & 1 & 0 & 0 & 0 & 0 & 0 & 0 & 0\\
        {\epsilon} & 0 & 0 & 1 & 1 & 0 & 0 & 0 & 0 & 0\\
        {m} & 0 & 0 & 0 & 0 & 1 & 0 & 0 & 0 & 0\\
        {e} & 0 & 0 & 0 & 0 & 1 & 0 & 0 & 0 & 0\\
        \chi & 0 & 0 & 0 & 0 & 0 & 1 & 1 & 0 & 0\\
        \bar\chi & 0 & 0 & 0 & 0 & 0 & 0 & 0 & 1 & 1
        \end{pNiceMatrix},\quad\quad\quad\quad
    B^\text{TC} \hs =\hs \begin{pNiceMatrix}[first-row,first-col]
        & 1 & \epsilon & m & e & \chi & \bar\chi\\
        1 & 1 & 0 & 0 & 0 & 0 & 0\\
        {\epsilon} & 0 & 1 & 0 & 0 & 0 & 0\\
        {m} & 0 & 0 & 1 & 0 & 0 & 0\\
        {e} & 0 & 0 & 0 & 1 & 0 & 0\\
    \end{pNiceMatrix},
    \end{equation*}
    \begin{equation*}
    B^\text{TC-DI} = B^\text{TC}B^\text{DI} \hs =\hs \begin{pNiceMatrix}[first-row,first-col]
        & 1\bar{1} & \psi\bar\psi & \psi\bar{1} & 1\bar{\psi} & \sigma\bar{\sigma} & \sigma\bar 1 & \sigma\bar\psi & 1\bar \sigma & \psi\bar\sigma\\
        1 & 1 & 1 & 0 & 0 & 0 & 0 & 0 & 0 & 0\\
        {\epsilon} & 0 & 0 & 1 & 1 & 0 & 0 & 0 & 0 & 0\\
        {m} & 0 & 0 & 0 & 0 & 1 & 0 & 0 & 0 & 0\\
        {e} & 0 & 0 & 0 & 0 & 1 & 0 & 0 & 0 & 0\\
    \end{pNiceMatrix}.
    \end{equation*}
    
    The branching matrix $B^\text{TC-DI}$ shows that there are $6$ interdomain excitation species:
    \begin{equation*}
        (1\bar 1,1),\quad\quad (\psi\bar\psi,1),\quad\quad (\psi\bar 1,\epsilon),\quad\quad (1\bar\psi,\epsilon),\quad\quad (\sigma\bar\sigma,m),\quad\quad (\sigma\bar\sigma,e).
    \end{equation*}
    whose quantum dimensions are all $1$. The braiding of the domain-wall quasiparticles and interdomain excitations are
    \begin{equation*}
    S^\text{DW} \hs =\hs \frac{1}{2}\quad \begin{pNiceMatrix}[first-row,first-col]
        & (1\bar{1},1) & (\psi\bar\psi,1) & (\psi\bar{1},\epsilon) & (1\bar{\psi},\epsilon) & (\sigma\bar{\sigma},m) & (\sigma\bar\sigma,e)\\
        1 & 1 & 1 & 1 & 1 & 1 & 1\\
        {\epsilon} & 1 & 1 & 1 & 1 & -1 & -1\\
        {m} & 1 & 1 & -1 & -1 & 1 & -1\\
        {e} & 1 & 1 & -1 & -1 & -1 & 1\\
        \chi & \sqrt{2} & -\sqrt{2} & -\sqrt{2} & \sqrt{2} & 0 & 0\\
        \bar\chi & \sqrt{2} & -\sqrt{2} & \sqrt{2} & -\sqrt{2} & 0 & 0\\
        \end{pNiceMatrix}.
    \end{equation*}
    
    Using the domain-wall Verlinde formulae, we find the nonzero independent fusion rules of interdomain excitations:
    \begin{align*}
    &(\psi\bar\psi, 1)\times(\psi\bar\psi, 1) = (\psi\bar 1, \epsilon)\times(\psi\bar 1, \epsilon) = (1\bar\psi, \epsilon)\times(1\bar\psi, \epsilon) = (1\bar 1,1),\\
    &(\psi\bar\psi, 1)\times(\psi\bar 1,\epsilon) = (1\bar\psi, \epsilon),\quad\quad\quad\quad (\psi\bar\psi, 1)\times(1\bar\psi,\epsilon) = (\psi\bar 1, \epsilon),\quad\quad\quad\quad (\psi\bar 1, \epsilon)\times(1\bar\psi,\epsilon) = (\psi\bar\psi, 1),\\
    &(\psi\bar\psi, 1) \times (\sigma\bar\sigma, e) = (\psi\bar 1, \epsilon) \times (\sigma\bar\sigma, m) = (1\bar\psi, \epsilon) \times (\sigma\bar\sigma, m) = (\sigma\bar\sigma, e),\\
    &(\psi\bar\psi, 1) \times (\sigma\bar\sigma, m) = (\psi\bar 1, \epsilon) \times (\sigma\bar\sigma, e) = (1\bar\psi, \epsilon) \times (\sigma\bar\sigma, e) = (\sigma\bar\sigma, m),\\
    &(\sigma\bar\sigma, e)\times (\sigma\bar\sigma, e) = (\sigma\bar\sigma, m)\times (\sigma\bar\sigma, m) = \frac{1}{2}(1\bar 1, 1) + \frac{1}{2}(\psi\bar\psi, 1),\\
    &(\sigma\bar\sigma, e)\times (\sigma\bar\sigma, m) = \frac{1}{2}(\psi\bar 1, \epsilon) + \frac{1}{2}(1\bar\psi, \epsilon),
    \end{align*}
    and the nonzero independent fusion rules of domain-wall quasiparticles:
    \begin{align*}
    &e\times e = m\times m = \epsilon\times\epsilon = 1,\quad\quad\quad\quad e\times m = \epsilon,\quad\quad\quad\quad e\times\epsilon = m,\quad\quad\quad\quad m\times\epsilon = e,\\
    &e\times\chi= m\times\chi = \epsilon\times\bar\chi =\bar\chi,\quad\quad\quad\quad e\times\bar\chi= m\times\bar\chi=\epsilon\times\chi = \chi,\\
    &\chi\times\chi=\bar\chi\times\bar\chi= 1+\epsilon,\quad\quad\quad\quad\quad\quad\chi\times\bar\chi= e + m.
    \end{align*}
    
    \subsection{The composite system of the $\text{su}(2)_{10}$ and $\text{so}(5)_1$ phases\label{appendix:suso}}
    
    There are $11$ anyon species in the $\text{su}(2)_{10}$ topological order, labeled by integers $0\le a\le 10$, where $0$ is the trivial anyon. The fusion tensor $N_{ab}^c = 1$ if and only if $(a + b + c)$ is an even number and $|a-b|\le c\le\min(a+b, 20-a-b)$; otherwise $N_{ab}^c = 0$. The quantum dimensions of anyons and the $S$-matrix elements are
    \begin{equation*}
        d_c = \frac{\sin\frac{a+1}{12}\pi}{\sin\frac{\pi}{12}},\quad\quad\quad\quad S_{ab}^{\text{su}(2)_{10}} = \sum_{c=0}^{10}\frac{N_{ab}^c\theta_cd_c}{\theta_a\theta_b},
    \end{equation*}
    where $\theta_c = e^{\frac{i\pi}{24}a(a+2)}$.
    
    Condensing anyon $6$ in the $\text{su}(2)_{10}$ phase results in the $\text{so}(5)_1$ phase. There are $3$ anyons in the $\text{so}(5)_1$ phase, labeled by $1, \psi, \sigma$, whose quantum dimensions are $d_1 = d_\psi = 1$, $d_\sigma = \sqrt{2}$. The $S$-matrix of the $\text{so}(5)_1$ phase is
    \begin{equation*}
        S^{\text{so}(5)_1} \hs =\hs \frac{1}{2}\quad \begin{pNiceMatrix}[first-row,first-col]
        & 1 & \psi & \sigma\\
        1 & 1 & 1 & \sqrt{2}\\
        \psi & 1 & 1 & -\sqrt{2}\\
        \sigma & \sqrt{2} & -\sqrt{2} & 0\\
        \end{pNiceMatrix}.
    \end{equation*}
    
    There are $6$ quasiparticle species in the gapped domain wall between the $\text{su}(2)_{10}$ and $\text{so}(5)_{1}$ topological orders, labeled by $1$, $\psi$, $\sigma$, $u$, $u\psi$, and $u\sigma$. The branching matrices are
    \begin{align*}
    \addtocounter{MaxMatrixCols}{12}
    B^{\text{su}(2)_{10}} \hs &=\hs \begin{pNiceMatrix}[first-row,first-col]
        & 0 & 1 & 2 & 3 & 4 & 5 & 6 & 7 & 8 & 9 & 10 \\
        1 & 1 & 0 & 0 & 0 & 0 & 0 & 1 & 0 & 0 & 0 & 0\\
        \psi & 0 & 0 & 0 & 0 & 1 & 0 & 0 & 0 & 0 & 0 & 1\\
        \sigma & 0 & 0 & 0 & 1 & 0 & 0 & 0 & 1 & 0 & 0 & 0\\
        u & 0 & 1 & 0 & 0 & 0 & 1 & 0 & 1 & 0 & 0 & 0\\
        u\psi & 0 & 0 & 0 & 1 & 0 & 1 & 0 & 0 & 0 & 1 & 0\\
        u\sigma & 0 & 0 & 1 & 0 & 1 & 0 & 1 & 0 & 1 & 0 & 0\\
    \end{pNiceMatrix},\quad\quad\quad\quad
    B^{\text{so}(5)_{1}} \hs =\hs \begin{pNiceMatrix}[first-row,first-col]
        & 1 & \sigma & \psi & u & v & w\\
        1 & 1 & 0 & 0 & 0 & 0 & 0\\
        \psi & 0 & 1 & 0 & 0 & 0 & 0\\
        \sigma & 0 & 0 & 1 & 0 & 0 & 0\\
    \end{pNiceMatrix},\\
    &B^{\text{so}(5)_1\text{-su}(2)_{10}} = B^{\text{so}(5)_1}B^{\text{su}(2)_{10}} \hs =\hs \begin{pNiceMatrix}[first-row,first-col]
        & 0 & 1 & 2 & 3 & 4 & 5 & 6 & 7 & 8 & 9 & 10 \\
        1 & 1 & 0 & 0 & 0 & 0 & 0 & 1 & 0 & 0 & 0 & 0\\
        \psi & 0 & 0 & 0 & 0 & 1 & 0 & 0 & 0 & 0 & 0 & 1\\
        \sigma & 0 & 0 & 0 & 1 & 0 & 0 & 0 & 1 & 0 & 0 & 0\\
    \end{pNiceMatrix},
    \end{align*}
    which shows that there are $6$ interdomain excitation species:
    \begin{equation*}
        (0, 1),\quad\quad (6, 1),\quad\quad (4, \psi),\quad\quad (10, \psi),\quad\quad (3, \sigma),\quad\quad (7, \sigma),
    \end{equation*}
    with $d_{(0,1)} = d_{(6, 1)} = d_{(4,\psi)} = d_{(10,\psi)} = 1$, and $d_{(3, \sigma)} = d_{(7, \sigma)} = \sqrt{2}$. The domain-wall $S$-matrix is
    \begin{equation*}
    S^\text{DW} \hs =\hs \frac{1}{2}\quad \begin{pNiceMatrix}[first-row,first-col]
        & (0,1) & (6,1) & (4,\psi) & (10,\psi) & (3,\sigma) & (7,\sigma)\\
        1 & 1 & 1 & 1 & 1 & \sqrt{2} & \sqrt{2}\\
        \psi  & 1 & 1 & 1 & 1 & -\sqrt{2} & -\sqrt{2}\\
        \sigma & \sqrt{2} & \sqrt{2} & -\sqrt{2} & -\sqrt{2} & 0 & 0\\
        u & \frac{\sqrt{2} +\sqrt{6}}{2} & \frac{\sqrt{2} -\sqrt{6}}{2} & \frac{-\sqrt{2} +\sqrt{6}}{2} & -\frac{\sqrt{2} +\sqrt{6}}{2} & \sqrt{2} & -\sqrt{2}\\
        u\psi & \frac{\sqrt{2} +\sqrt{6}}{2} & \frac{\sqrt{2} -\sqrt{6}}{2} & \frac{-\sqrt{2} +\sqrt{6}}{2} & -\frac{\sqrt{2} +\sqrt{6}}{2} & -\sqrt{2} & \sqrt{2}\\
        u\sigma & (1+\sqrt{3}) & (1-\sqrt{3}) & (1-\sqrt{3}) & (1+\sqrt{3}) & 0 & 0\\
        \end{pNiceMatrix}.
    \end{equation*}
    
    Using the domain-wall Verlinde formulae, we obtain the fusion rules of the interdomain excitations and domain-wall quasiparticles:
    \begin{align*}
        &\psi\times\psi= 1,\quad\quad\ \ \sigma\times\sigma= 1+\psi,\quad\quad\ \ \sigma\times\psi=\sigma,\\
        &u\times\psi = u\psi,\quad\quad u\psi\times \psi = u,\quad\quad\quad\ \  u\sigma\times\sigma= u + u\psi,\quad\quad u\times \sigma = u\psi\times\sigma = u\sigma\times \psi = u\sigma,\\
        &u\times u= u\psi\times u\psi = 1 + u\sigma,\quad\quad\quad\quad\ \ u\times u\psi = \psi + u\sigma,\\
        &u\times u\sigma = u\psi\times u\sigma = \sigma + u + u\psi,\quad\quad u\sigma\times u\sigma = 1 + \psi + 2u\sigma,\\ \\
        &(6, 1)\times(10,\psi) = (4,\psi),\quad\quad (4,\psi)\times(10,\psi) = (6,1),\quad\quad (10,\psi)\times(10,\psi) = (0, 1),\nonumber\\
        &(4,\psi)\times(4,\psi) = (6,1)\times(6, 1) = (2 - \sqrt{3})(0, 1) + (\sqrt{3} - 1)(6, 1),\nonumber\\
        &(4,\psi)\times(6, 1) = (2 - \sqrt{3})(10,\psi) + (\sqrt{3} - 1)(4,\psi),\nonumber\\
        &(6, 1)\times(7,\sigma) = (4,\psi)\times(3,\sigma) = \frac{3 - \sqrt{3}}{2}(3,\sigma) + \frac{\sqrt{3} - 1}{2}(7,\sigma),\nonumber\\
        &(6, 1)\times(3,\sigma) = (4,\psi)\times(7,\sigma) = \frac{\sqrt{3} - 1}{2}(3,\sigma) + \frac{3 - \sqrt{3}}{2}(7,\sigma),\nonumber\\
        &(10, \psi)\times(3,\sigma) = (7,\sigma),\quad\quad (10, \psi)\times(7,\sigma) = (3, \sigma),\nonumber\\
        &(3,\sigma)\times(3,\sigma) = (7,\sigma)\times(7,\sigma) = \left(1 - \frac{\sqrt{3}}{3}\right)(0, 1) + \frac{\sqrt{3}}{3}(6, 1) + (4,\psi),\nonumber\\
        &(3,\sigma)\times(7,\sigma) = \left(1 - \frac{\sqrt{3}}{3}\right)(10, \psi) + \frac{\sqrt{3}}{3}(4, \psi) + (6,1).
    \end{align*}
    
    \subsection{The composite system of the $D(S_3)$ and $\Z_2$ toric code phases}
    
    There are $8$ anyon species in the $D(S_3)$ topological order, labeled by $A$, $B$, $C$, $D$, $E$, $F$, $G$, and $H$, whose quantum dimensions are $d_A = d_B = 1, d_C = d_F = d_G = d_H = 2, d_D = d_E = 3$. Here $A$ is the trivial anyon. The $S$-matrix is
    \begin{equation*}
    S^{D(S_3)} \hs = \hs \frac{1}{6}\quad{ \begin{pNiceMatrix}[first-row,first-col]
        & A & B & C & D & E & F & G & H \\
        A & 1 & 1 & 2 & 3 & 3 & 2 & 2 & 2 \\
        B & 1 & 1 & 2 & -3 & -3 & 2 & 2 & 2 \\
        C & 2 & 2 & 4 & 0 & 0 & -2 & -2 & -2 \\
        D & 3 & -3 & 0 & 3 & -3 & 0 & 0 & 0 \\
        E & 3 & -3 & 0 & -3 & 3 & 0 & 0 & 0 \\
        F & 2 & 2 & -2 & 0 & 0 & 4 & -2 & -2 \\
        G & 2 & 2 & -2 & 0 & 0 & -2 & -2 & 4 \\
        H & 2 & 2 & -2 & 0 & 0 & -2 & 4 & -2 \\
        \end{pNiceMatrix}}.
    \end{equation*}
    
    Condensing anyon $C$ in the $D(S_3)$ topological order results in $\Z_2$ toric code phase. There are $6$ species of quasiparticles in the gapped domain wall between the $D(S_3)$ and $\Z_2$ topological orders, labeled by $1$, $e$, $m$, $\epsilon$, $\sigma$, and $\tau$. The quantum dimensions of domain-wall quasiparticles are $d_1 = d_\epsilon = d_e = d_m = 1$, $d_\sigma = 2$, and $d_\tau = 3$. The branching matrices are
    \begin{equation*}
    B^{D(S_3)} \hs =\hs \begin{pNiceMatrix}[first-row,first-col]
        & A & B & C & D & E & F & G & H \\
        1 & 1 & 0 & 1 & 0 & 0 & 0 & 0 & 0\\
        e & 0 & 1 & 1 & 0 & 0 & 0 & 0 & 0\\
        m & 0 & 0 & 0 & 1 & 0 & 0 & 0 & 0\\
        \epsilon & 0 & 0 & 0 & 0 & 1 & 0 & 0 & 0\\
        \sigma & 0 & 0 & 0 & 1 & 1 & 0 & 0 & 0\\
        \tau & 0 & 0 & 0 & 0 & 0 & 1 & 1 & 1
        \end{pNiceMatrix},\quad\quad\quad\quad
    B^\text{TC} \hs =\hs \begin{pNiceMatrix}[first-row,first-col]
        & 1 & e & m & \epsilon & \sigma & \tau\\
        1 & 1 & 0 & 0 & 0 & 0 & 0\\
        e & 0 & 1 & 0 & 0 & 0 & 0\\
        m & 0 & 0 & 1 & 0 & 0 & 0\\
        \epsilon & 0 & 0 & 0 & 1 & 0 & 0\\
    \end{pNiceMatrix},
    \end{equation*}
    \begin{equation*}
    B = B^\text{TC}B^{D(S_3)} \hs =\hs\begin{pNiceMatrix}[first-row,first-col]
        & A & B & C & D & E & F & G & H \\
        1 & 1 & 0 & 1 & 0 & 0 & 0 & 0 & 0\\
        e & 0 & 1 & 1 & 0 & 0 & 0 & 0 & 0\\
        m & 0 & 0 & 0 & 1 & 0 & 0 & 0 & 0\\
        \epsilon & 0 & 0 & 0 & 0 & 1 & 0 & 0 & 0\\
    \end{pNiceMatrix},
    \end{equation*}
    indicating that there are $6$ species of interdomain excitations:
    \begin{equation*}
        (A, 1),\quad\quad (C, 1),\quad\quad (B, e),\quad\quad (C, e),\quad\quad (D, m),\quad\quad (E, \epsilon),
    \end{equation*}
    whose quantum dimensions are all $1$. The domain-wall $S$-matrix is
    \begin{equation*}
    S^\text{DW} \hs =\hs \frac{1}{2}\quad \begin{pNiceMatrix}[first-row,first-col]
        & (A,1) & (C,1) & (B,e) & (C,e) & (D,m) & (E,\epsilon)\\
        1 & 1 & 1 & 1 & 1 & 1 & 1\\
        e & 1 & 1 & 1 & 1 & -1 & -1\\
        m & 1 & 1 & -1 & -1 & 1 & -1\\
        \epsilon & 1 & 1 & -1 & -1 & -1 & 1\\
        \sigma & 2 & -1 & -2 & 1 & 0 & 0\\
        \tau & 2 & -1 & 2 & -1 & 0 & 0\\
        \end{pNiceMatrix}.
    \end{equation*}
    
    The fusion rules of the interdomain excitations and domain-wall quasiparticles are:
    \begin{align*}
        &e\times e = m\times m = \epsilon\times\epsilon= 1,\quad\quad e\times m = \epsilon,\quad\quad e\times\epsilon = m,\quad\quad m\times\epsilon = e,\nonumber\\
        &e\times\sigma = m\times\tau = \epsilon\times\tau=\sigma,\quad\quad e\times\tau = m\times\sigma = \epsilon\times\sigma = \tau,\nonumber\\
        &\tau\times\tau = \sigma\times\sigma = 1+ e+\tau,\quad\quad \tau\times\sigma= m+\epsilon+\sigma,\\
        &(C, 1)\times(C, 1) = (C, e)\times(C, e) = \frac{1}{2}(A, 1) + \frac{1}{2}(C, 1),\\
        &(C, 1)\times(C, e) = \frac{1}{2}(B, e) + \frac{1}{2}(C, e),\\
        &(C, 1)\times(B, e) = (C, e),\quad\quad(C, e)\times(B, e) = (C, 1),\quad\quad(B, e)\times(B, e) = (A, 1),\\
        &(C, 1)\times(D, m) = (B, e)\times(E,\epsilon) = (C, e)\times(E,\epsilon) = (D, m),\\
        &(C, 1)\times(E,\epsilon) = (B, e)\times(D, m) = (C, e)\times(D, m) = (E, \epsilon),\\
        &(D, m)\times(D, m) = (E, \epsilon)\times(E, \epsilon) = \frac{1}{3}(A, 1) + \frac{2}{3}(C, 1),\\
        &(D, m)\times(E, \epsilon) = \frac{1}{3}(B, e) + \frac{2}{3}(C, e).
    \end{align*}
    \end{widetext}
    
    \bibliography{StringNet}% Produces the bibliography via BibTeX.
    \end{document}

%% file: totalinput.tex
\newcommand{\FigureDomainWall}{
\begin{tikzpicture}[baseline={([yshift=-.8ex]current bounding box.center)},scale=.8]
    \filldraw [red,opacity=0.2] (1.8,7.4) -- (-0.2,5.4) -- (2.4,5.4) -- (4.4,7.4);
    \filldraw [black,opacity=0.2] (4.4,7.4) -- (2.4,5.4) -- (2.8,5.4) -- (4.8,7.4);
    \filldraw [blue,opacity=0.2] (4.8,7.4) -- (2.8,5.4) -- (5.4,5.4) -- (7.4,7.4);
    \node [centered, rotate=0.] at (2.8,7.6) {\scriptsize {Domain $\mathcal{A}$}};
    \node [centered, rotate=0.] at (6.4,7.6) {\scriptsize {Domain $\mathcal{B}$}};
    \node [centered, rotate=0.] at (4.6,7.9) {\scriptsize {Domain}};
    \node [centered, rotate=0.] at (4.6,7.6) {\scriptsize {Wall}};

    \draw [very thick] (1.6,6.4) ..controls (1.804,6.52) and (2.009,6.64) .. (2.2,6.7)
    ..controls (2.391,6.76) and (2.569,6.76) .. (2.8,6.7)
    ..controls (3.031,6.64) and (3.316,6.52) .. (3.6,6.4)
    ..controls (3.884,6.28) and (4.169,6.16) .. (4.4,6.1)
    ..controls (4.631,6.04) and (4.809,6.04) .. (5.,6.1)
    ..controls (5.191,6.16) and (5.396,6.28) .. (5.6,6.4);
    \fill [red] (1.6,6.4) circle (0.1);
    \fill [blue] (5.6,6.4) circle (0.1);
    \node [centered, rotate=0.] at (1.3,6.2) {\scriptsize $a$};
    \node [centered, rotate=0.] at (5.85,6.55) {\scriptsize $r$};

    \fill [] (3.95,6.75) circle (0.1);
    \node [centered, rotate=0.] at (4.15,6.95) {\scriptsize $\alpha$};

    \draw[ultra thick, ->] (-0.4, 5.2) -- (6.0,5.2);
    \node [centered, rotate=0.] at (2.8,4.9) {\scriptsize {Space}};
\end{tikzpicture}
}

\newcommand{\FigureBraiding}{
\hs\begin{tikzpicture}[baseline={([yshift=-.8ex]current bounding box.center)},scale=0.45]
    \draw [white] (2, 4.1) -- (3., 4.1);
    \node [centered, rotate=0.] at (-0.9,7.6) {\normalsize $S^\text{DW}_{\alpha(a,r)}$};
    \node [centered, rotate=0.] at (-0.97,6.2) {\normalsize $S^\text{DW}_{1(1,1)}$};
    \draw [] (-2.2, 6.9) -- (0.3, 6.9);
    \node [centered, rotate=0.] at (.9,6.9) {\normalsize $:=$};

    \filldraw [red,opacity=0.2](1.6,8.8)--(1.6,4.8)--(3.4,4.8)--(3.4,8.8);
    \filldraw [black,opacity=0.2](4.6,8.8)--(4.6,4.8)--(3.4,4.8)--(3.4,8.8);
    \filldraw [blue,opacity=0.2](4.6,8.8)--(4.6,4.8)--(6.4,4.8)--(6.4,8.8);

    \draw [rounded corners,very thick] (3.8,6.8) -- (5.6,6.8) -- (5.6,8.) -- (2.4,8.) -- (2.4, 6.8) -- (3.4,6.8);
    \node [centered, rotate=0.] at (4.4,5.4) {\scriptsize $\alpha$};

    \draw [rounded corners,very thick] (4.4, 7.) -- (4.4,7.6) -- (3.6,7.6) -- (3.6,5.6) -- (4.4,5.6) -- (4.4,6.6);
    \node [centered, rotate=0.] at (2.0,7.8) {\scriptsize $a$};
    \node [centered, rotate=0.] at (6.0,7.8) {\scriptsize $r$};
\end{tikzpicture}\hs
}

\newcommand{\FigureVerlindeA}{
\begin{tikzpicture}[baseline={([yshift=-.8ex]current bounding box.center)},scale=.5]
    \fill [red,opacity=0.2] (1.8,4.5) -- (3.6,4.5) -- (3.6,7.4) -- (1.8,7.4);
    \fill [black,opacity=0.2] (3.6,4.5) -- (4.4,4.5) -- (4.4,7.4) -- (3.6,7.4);
    \fill [blue,opacity=0.2] (4.4,4.5) -- (6.2,4.5) -- (6.2,7.4) -- (4.4,7.4);
    
    \draw [rounded corners, very thick] (4.05,6.6) -- (2.4,6.6) -- (2.4,4.8) -- (5.6,4.8) -- (5.6,6.6) -- (4.35,6.6);
    \node [centered, rotate=0.] at (2.1,5.65) {\scriptsize $a$};
    \node [centered, rotate=0.] at (5.9,5.65) {\scriptsize $r$};
    \draw [rounded corners, very thick] (4.05,6.1) -- (2.9,6.1) -- (2.9,5.3) -- (5.1,5.3) -- (5.1,6.1) -- (4.35,6.1);
    \node [centered, rotate=0.] at (3.2,5.65) {\scriptsize $b$};
    \node [centered, rotate=0.] at (4.8,5.65) {\scriptsize $s$};

    \draw [rounded corners, very thick] (3.8, 5.95) -- (3.8,5.6) -- (4.2,5.6) -- (4.2,7.1) -- (3.8,7.1) -- (3.8,6.75);
    \draw [very thick] (3.8, 6.25) -- (3.8,6.45);
    \node [centered, rotate=0.] at (4.4,7.1) {\scriptsize $\alpha$};

    \node [centered, rotate=0.] at (4.0,3.8) {\scriptsize {(a)}};
\end{tikzpicture}
}

\newcommand{\FigureVerlindeB}{
\begin{tikzpicture}[baseline={([yshift=-.8ex]current bounding box.center)},scale=.5]
    \fill [red,opacity=0.2] (1.8,4.5) -- (3.6,4.5) -- (3.6,7.4) -- (1.8,7.4);
    \fill [black,opacity=0.2] (3.6,4.5) -- (4.4,4.5) -- (4.4,7.4) -- (3.6,7.4);
    \fill [blue,opacity=0.2] (4.4,4.5) -- (6.2,4.5) -- (6.2,7.4) -- (4.4,7.4);
    
    \draw [rounded corners, very thick] (4.05,6.6) -- (2.4,6.6) -- (2.4,4.8) -- (5.6,4.8) -- (5.6,6.6) -- (4.35,6.6);
    \node [centered, rotate=0.] at (2.1,5.65) {\scriptsize $c$};
    \node [centered, rotate=0.] at (5.9,5.65) {\scriptsize $t$};
    \draw [rounded corners, very thick, dashed] (4.05,6.1) -- (2.9,6.1) -- (2.9,5.3) -- (5.1,5.3) -- (5.1,6.1) -- (4.35,6.1);
    \node [centered, rotate=0.] at (3.2,5.65) {\scriptsize $1$};
    \node [centered, rotate=0.] at (4.8,5.65) {\scriptsize $1$};

    \draw [rounded corners, very thick] (3.8, 5.95) -- (3.8,5.6) -- (4.2,5.6) -- (4.2,7.1) -- (3.8,7.1) -- (3.8,6.75);
    \draw [very thick] (3.8, 6.25) -- (3.8,6.45);
    \node [centered, rotate=0.] at (4.4,7.1) {\scriptsize $\alpha$};

    \node [centered, rotate=0.] at (4.0,3.8) {\scriptsize {(b)}};
    
    \draw [] (-3.7, 5.9) -- (-4.2, 5.9);
    \draw [] (-3.7, 6.1) -- (-4.2, 6.1);
    \node [centered, rotate=0.] at (-2.4,6.0) {\Huge $\sum$};
    \node [centered, rotate=0.] at (-3.1,5.4) {\tiny $(c,t)$};
    \node [centered, rotate=0.] at (-2.42,5.4) {\tiny $\in$};
    \node [centered, rotate=0.] at (-1.8,5.37) {\tiny $\mathcal{L}_\text{ID}$};
    \node [centered, rotate=0.] at (0.1,6.05) {\normalsize $N_{(a,r)(b,s)}^{(c,t)}$};
\end{tikzpicture}
}

\newcommand{\FigureVerlindeC}{
\begin{tikzpicture}[baseline={([yshift=-.8ex]current bounding box.center)},scale=.5]
    \fill [red,opacity=0.2] (1.8,5.7) -- (3.2,5.7) -- (3.2,8.1) -- (1.8,8.1);
    \fill [black,opacity=0.2] (3.2,5.7) -- (4.8,5.7) -- (4.8,8.1) -- (3.2,8.1);
    \fill [blue,opacity=0.2] (4.8,5.7) -- (6.2,5.7) -- (6.2,8.1) -- (4.8,8.1);

    \draw [rounded corners,very thick] (3.7,7.2) -- (2.4,7.2) -- (2.4,6.1) -- (5.6,6.1) -- (5.6,7.2) -- (4.6,7.2);
    \draw [very thick] (3.9,7.2) -- (4.4,7.2);
    \node [centered, rotate=0.] at (2.1,6.6) {\scriptsize $a$};
    \node [centered, rotate=0.] at (5.85,6.6) {\scriptsize $r$};

    \draw [rounded corners,very thick] (3.4,7.0) -- (3.4,6.6) -- (3.8,6.6) -- (3.8,7.8) -- (3.4,7.8) -- (3.4,7.4);
    \node [centered, rotate=0.] at (3.25,6.5) {\scriptsize $\alpha$};
    \draw [rounded corners,very thick] (4.1,7.0) -- (4.1,6.6) -- (4.5,6.6) -- (4.5,7.8) -- (4.1,7.8) -- (4.1,7.4);
    \node [centered, rotate=0.] at (4.6,6.5) {\scriptsize $\beta$};

    \node [centered, rotate=0.] at (4.0,5.0) {\scriptsize {(c)}};
    \draw[white] (4.0, 8.2) -- (4.8, 8.2);
\end{tikzpicture}
}

\newcommand{\FigureVerlindeD}{
\begin{tikzpicture}[baseline={([yshift=-.8ex]current bounding box.center)},scale=.5]
    \fill [red,opacity=0.2] (1.8,5.7) -- (3.2,5.7) -- (3.2,8.1) -- (1.8,8.1);
    \fill [black,opacity=0.2] (3.2,5.7) -- (4.8,5.7) -- (4.8,8.1) -- (3.2,8.1);
    \fill [blue,opacity=0.2] (4.8,5.7) -- (6.2,5.7) -- (6.2,8.1) -- (4.8,8.1);

    \draw [rounded corners,very thick] (3.7,7.2) -- (2.4,7.2) -- (2.4,6.1) -- (5.6,6.1) -- (5.6,7.2) -- (4.6,7.2);
    \draw [very thick] (3.9,7.2) -- (4.4,7.2);
    \node [centered, rotate=0.] at (2.1,6.6) {\scriptsize $a$};
    \node [centered, rotate=0.] at (5.85,6.6) {\scriptsize $r$};

    \draw [rounded corners,very thick] (3.4,7.0) -- (3.4,6.6) -- (3.8,6.6) -- (3.8,7.8) -- (3.4,7.8) -- (3.4,7.4);
    \node [centered, rotate=0.] at (3.25,6.5) {\scriptsize $\gamma$};
    \draw [rounded corners,very thick, dashed] (4.1,7.0) -- (4.1,6.6) -- (4.5,6.6) -- (4.5,7.8) -- (4.1,7.8) -- (4.1,7.4);
    \node [centered, rotate=0.] at (4.6,6.5) {\scriptsize $1$};

    \node [centered, rotate=0.] at (4.0,5.0) {\scriptsize {(d)}};
    \draw [white] (-4.1, 8.2) -- (-4.1, 8.2);
    
    \draw [] (-2.6, 6.8) -- (-3.1, 6.8);
    \draw [] (-2.6, 7.0) -- (-3.1, 7.0);
    \node [centered, rotate=0.] at (-0.8,6.9) {\Huge $\sum$};
    \node [centered, rotate=0.] at (-1.5,6.3) {\tiny $\gamma$};
    \node [centered, rotate=0.] at (-1.1,6.3) {\tiny $\in$};
    \node [centered, rotate=0.] at (-0.3,6.27) {\tiny $\mathcal{L}_\text{DW}$};
    \node [centered, rotate=0.] at (0.85,6.9) {\normalsize $N_{\alpha\beta}^{\gamma}$};
\end{tikzpicture}
}

\newcommand{\FigureLoopOperator}{
\begin{tikzpicture}[baseline={([yshift=-.8ex]current bounding box.center)},scale=1.2]
\draw [ultra thick, red] (0.0, 0.0) -- (1.4, 0.0);
\draw [ultra thick, gray] (1.4, 0.0) -- (1.8, 0.0);
\draw [ultra thick, blue] (1.8, 0.0) -- (3.2, 0.0);
\draw [ultra thick, red] (0.0, 0.04) -- (1.4, 0.04);
\draw [ultra thick, gray] (1.4, 0.04) -- (1.8, 0.04);
\draw [ultra thick, blue] (1.8, 0.04) -- (3.2, 0.04);
\node [centered, rotate=0.] at (0.7, -0.2) {\scriptsize Domain $\mathcal{A}$};
\node [centered, rotate=0.] at (2.5, -0.2) {\scriptsize Domain $\mathcal{B}$};
\draw [white] (4.0, 1.0) -- (4.0, 2.0);

\draw [thick, ->] (0.0, -0.4) -- (0.0, 2.2);
\node [left, rotate=0.] at (0., 0.38) {\scriptsize $\left|\psi\right\rangle$};
\node [left, rotate=0.] at (0., 1.) {\scriptsize $W_L^{(a,r)}\left|\psi\right\rangle$};
\node [left, rotate=0.] at (0., 1.65) {\scriptsize $C_L^{(a,r)}\left|\psi\right\rangle$};
\node [centered, rotate=0.] at (0.4, 2.0) {\scriptsize Time};

\draw [ultra thick] (1.6,1.) ellipse [x radius=0.8, y radius=0.6];
\draw [dashed] (0.0, 0.38) -- (3., 0.38);
\draw [dashed] (0.0, 1.) -- (3., 1.);
\draw [dashed] (0.0, 1.62) -- (3., 1.62);
\fill [red] (0.8, 1.) circle (0.08);
\fill [blue] (2.4, 1.) circle (0.08);
\node [centered, rotate=0.] at (0.95,1.1) {\scriptsize $a$};
\node [centered, rotate=0.] at (2.25,1.1) {\scriptsize $r$};
\node [centered, rotate=0.] at (1.3,0.6) {\scriptsize $L$};

\node [centered, rotate=0.] at (2.7, 0.6) {\scriptsize $W_L^{(a,r)}$};
\node [centered, rotate=0.] at (2.85, 1.4) {\scriptsize $\left(W_L^{(a,r)}\right)^\dagger$};
\end{tikzpicture}
}

\newcommand{\FigureSingleA}{
\begin{tikzpicture}[baseline={([yshift=-.8ex]current bounding box.center)},scale=0.4]
    \draw [white] (2.0, 7.) -- (3.0, 7.);
    \filldraw [red,opacity=0.2] (0.8,12.) -- (6.4,12.) -- (6.4,7.2) -- (0.8,7.2);
    
    \draw [very thick] (2.4,8.) ..controls (3.2,8.178) and (4.,8.356) .. (4.,8.8) ..controls (4.,9.244) and (3.2,9.956) .. (3.2,10.4) ..controls (3.2,10.844) and (4.,11.022) .. (4.8,11.2);
    \fill [red] (2.4,8.) circle (0.15);
    \fill [red] (4.8,11.2) circle (0.15);
    \node [centered, rotate=0.] at (1.85,8.) {\scriptsize $a^*$};
    \node [centered, rotate=0.] at (5.35,11.25) {\scriptsize $a$};
    \node [centered, rotate=0.] at (4.1,9.7) {\scriptsize $L$};
\end{tikzpicture}
}

\newcommand{\FigureSingleB}{
\begin{tikzpicture}[baseline={([yshift=-.8ex]current bounding box.center)},scale=.4]
    \draw [white] (2.0, 7.) -- (3.0, 6.8);
    \filldraw [red,opacity=0.2] (0.8,12.) -- (6.4,12.) -- (6.4,7.2) -- (0.8,7.2);
    
    \draw [very thick] (1.2,9.6) ..controls (1.471,9.751) and (1.742,9.902) .. (2.,9.8) ..controls (2.258,9.698) and (2.502,9.342) .. (2.8,9.2) ..controls (3.098,9.058) and (3.449,9.129) .. (3.8,9.2);
    \draw [very thick] (5.8,7.6) ..controls (5.489,7.689) and (5.178,7.778) .. (5.,8.) ..controls (4.822,8.222) and (4.778,8.578) .. (4.6,8.8) ..controls (4.422,9.022) and (4.111,9.111) .. (3.8,9.2);
    \draw [very thick] (3.8,9.2) ..controls (3.804,9.493) and (3.809,9.787) .. (4.,10.) ..controls (4.191,10.213) and (4.569,10.347) .. (4.8,10.6) ..controls (5.031,10.853) and (5.116,11.227) .. (5.2,11.6);
    \fill [red] (1.2,9.6) circle (0.15);
    \fill [red] (5.8,7.6) circle (0.15);
    \fill [red] (5.2,11.6) circle (0.15);
    \node [centered, rotate=0.] at (1.3,9.15) {\scriptsize $a^*$};
    \node [centered, rotate=0.] at (6.1,8.05) {\scriptsize $b^*$};
    \node [centered, rotate=0.] at (4.75,11.6) {\scriptsize $c$};
    \node [centered, rotate=0.] at (3.8,8.8) {\scriptsize $\mu$};
\end{tikzpicture}
}

\newcommand{\FigureSingleC}{
\begin{tikzpicture}[baseline={([yshift=-.8ex]current bounding box.center)},scale=.3]
    \draw [white] (2.0, 7.) -- (3.0, 6.8);
    \filldraw [red,opacity=0.2] (0.8,12.) -- (6.4,12.) -- (6.4,7.2) -- (0.8,7.2);

    \draw [rounded corners,very thick] (1.8, 9.1) -- (1.8, 7.6) -- (4.2,7.6) -- (4.2,10.6) -- (3.2,10.6);
    \draw [rounded corners,very thick] (2.8,10.6) -- (1.8,10.6)-- (1.8, 9.1);
    \draw [rounded corners,very thick] (5.3,10.1) -- (5.3, 8.6) -- (4.4,8.6);
    \draw [rounded corners,very thick] (4.,8.6) -- (3.,8.6) -- (3.,11.6) -- (5.3,11.6) -- (5.3,10.1);
    \node [centered, rotate=0.] at (1.4,8.8) {\scriptsize $a$};
    \node [centered, rotate=0.] at (5.75,10.4) {\scriptsize $b$};
\end{tikzpicture}
}

\newcommand{\FigureCorrespondenceA}{
\begin{tikzpicture}[baseline={([yshift=-.8ex]current bounding box.center)},scale=0.65]
    \filldraw [red,opacity=0.2] (1.6,6.8) -- (6.0,6.8) -- (7.2,8.0) -- (2.8,8.0);
    \fill [red] (5.6,7.2) circle (0.1);
    \node [centered, rotate=0.] at (5.9,7.3) {\scriptsize $a$};
    \node [centered, rotate=0.] at (3.9,7.5) {\scriptsize {Parent}};
    \node [centered, rotate=0.] at (3.9,7.1) {\scriptsize {phase $\mathcal{A}$}};

    \filldraw [black,opacity=0.2] (1.6,5.8) -- (6.0,5.8) -- (7.2,7.0) -- (6.2,7.0) -- (6.0,6.8) -- (2.6,6.8);
    \fill [black] (5.6,6.2) circle (0.1);
    \node [centered, rotate=0.] at (5.9,6.3) {\scriptsize $\alpha$};
    \node [centered, rotate=0.] at (3.9,6.5) {\scriptsize {Intermediate}};
    \node [centered, rotate=0.] at (3.9,6.1) {\scriptsize {phase $\mathcal{T}$}};

    \filldraw [blue,opacity=0.2] (1.6,4.8) -- (6.0,4.8) -- (7.2,6.0) -- (6.2,6.0) -- (6.0,5.8) -- (2.6,5.8);
    \node [centered, rotate=0.] at (5.9,5.3) {\scriptsize $r$};
    \node [centered, rotate=0.] at (3.9,5.5) {\scriptsize {Child}};
    \node [centered, rotate=0.] at (3.9,5.1) {\scriptsize {phase $\mathcal{B}$}};

    \fill [blue] (5.6,5.2) circle (0.1);
    \draw [ultra thick, ->] (6.6,8.3) -- (6.6,4.7);
    \node [centered, rotate=0.] at (7.2,7.3) {\scriptsize {Time}};
    \draw [white] (4.0,4.2) -- (5.0,4.2);
    \draw [white] (4.0,8.4) -- (5.0,8.4);
\end{tikzpicture}
}

\newcommand{\FigureCorrespondenceB}{
\begin{tikzpicture}[baseline={([yshift=-.8ex]current bounding box.center)},scale=1]
    \draw [] (-0.1,1.0) -- (-0.1,2.0);
    \draw [] (1.1,1.0) -- (1.3,1.5);
    \draw [] (1.3,1.5) -- (1.1,2.0);
    \draw [thick] (0.5,1.1) -- (0.5,1.6);
    \draw [thick] (0.5,1.6) -- (0.1,1.8);
    \draw [thick] (0.5,1.6) -- (0.9,1.8);
    \fill [blue] (0.1, 1.8) circle (0.05);
    \fill [blue] (0.9, 1.8) circle (0.05);
    \fill [blue] (0.5, 1.1) circle (0.05);
    \node [centered, rotate=0.] at (0.0,1.9) {\scriptsize $r$};
    \node [centered, rotate=0.] at (1.0,1.9) {\scriptsize $s$};
    \node [centered, rotate=0.] at (0.65,1.15) {\scriptsize $t$};
    \node [centered, rotate=0.] at (0.5,1.8) {\scriptsize $\mu$};
    \node [centered, rotate=0.] at (0.5,0.7) {\tiny {Child Excitation State}};

    \draw [ultra thick, ->] (1.5, 1.5) -- (3.0, 1.5);
    \node [centered, rotate=0.] at (2.15,1.7) {\tiny {Embedding}};
    \node [centered, rotate=0.] at (3.5,1.5) {\Huge $\sum$};
    \node [centered, rotate=0.] at (3.2,1.2) {\tiny $abc$};
    \node [centered, rotate=0.] at (3.5,1.2) {\tiny $\in$};
    \node [centered, rotate=0.] at (3.8,1.2) {\tiny $\mathcal{L}_{\mathcal{A}}$};
    \node [centered, rotate=0.] at (3.0,0.9) {\tiny $1$};
    \node [centered, rotate=0.] at (3.2,0.9) {\tiny $\le$};
    \node [centered, rotate=0.] at (3.4,0.9) {\tiny $\nu$};
    \node [centered, rotate=0.] at (3.6,0.9) {\tiny $\le$};
    \node [centered, rotate=0.] at (3.95,0.9) {\tiny $N_{ab}^c$};

    \node [centered, rotate=0.] at (4.9,1.5) {\scriptsize $\vlc{r}{s}{t}{a}{b}{c}{\mu}{\nu}$};
    \draw [] (5.9,1.0) -- (5.9,2.0);
    \draw [] (7.1,1.0) -- (7.3,1.5);
    \draw [] (7.3,1.5) -- (7.1,2.0);
    \draw [thick] (6.5,1.1) -- (6.5,1.6);
    \draw [thick] (6.5,1.6) -- (6.1,1.8);
    \draw [thick] (6.5,1.6) -- (6.9,1.8);
    \fill [red] (6.1, 1.8) circle (0.05);
    \fill [red] (6.9, 1.8) circle (0.05);
    \fill [red] (6.5, 1.1) circle (0.05);
    \node [centered, rotate=0.] at (6.0,1.9) {\scriptsize $a$};
    \node [centered, rotate=0.] at (7.0,1.9) {\scriptsize $b$};
    \node [centered, rotate=0.] at (6.65,1.15) {\scriptsize $c$};
    \node [centered, rotate=0.] at (6.5,1.8) {\scriptsize $\nu$};
    \node [centered, rotate=0.] at (6.5,0.7) {\tiny {Parent Excitation States}};

    \draw [white] (4.0, 0.0) -- (5.0, 0.0);
    \draw [white] (4.0, 2.73) -- (5.0, 2.73);

\end{tikzpicture}
}

\newcommand{\FigureCorrespondenceC}{
\begin{tikzpicture}[baseline={([yshift=-.8ex]current bounding box.center)},scale=0.65]
    \filldraw [red,opacity=0.2] (1.6,4.8) -- (4.0,4.8) -- (4.0,8.) -- (1.6,8.);
    \filldraw [black,opacity=0.2] (4.0,4.8) -- (4.8,4.8) -- (4.8,8.) -- (4.0,8.);
    \filldraw [blue,opacity=0.2] (4.8,4.8) -- (7.2,4.8) -- (7.2,8.) -- (4.8,8.);
    \node [centered, rotate=0.] at (2.7,4.4) {\scriptsize {Domain $\mathcal{A}$}};
    \node [centered, rotate=0.] at (6.1,4.4) {\scriptsize {Domain $\mathcal{B}$}};

    \draw [very thick] (2.,7.2) ..controls (2.133,6.956) and (2.267,6.711) .. (2.6,6.8) ..controls (2.933,6.889) and (3.467,7.311) .. (3.8,7.4) ..controls (4.133,7.489) and (4.267,7.244) .. (4.4,7.);
    \fill [red] (2.,7.2) circle (0.1);
    \node [centered, rotate=0.] at (2.,7.5) {\scriptsize $a$};
    \fill [black] (4.4,7.0) circle (0.1);
    \node [centered, rotate=0.] at (4.7,7.0) {\scriptsize $\alpha$};

    \draw [very thick] (4.4,5.8) ..controls (4.533,5.533) and (4.667,5.267) .. (5.,5.4) ..controls (5.333,5.533) and (5.867,6.067) .. (6.2,6.2) ..controls (6.533,6.333) and (6.667,6.067) .. (6.8,5.8);
    \fill [black] (4.4,5.8) circle (0.1);
    \node [centered, rotate=0.] at (4.1,6.0) {\scriptsize $\alpha$};
    \fill [blue] (6.8,5.8) circle (0.1);
    \node [centered, rotate=0.] at (7.0,5.6) {\scriptsize $r$};

    \draw [white] (4.0,4.2) -- (5.0,4.2);
    \draw [white] (4.0,8.4) -- (5.0,8.4);
\end{tikzpicture}
}

\newcommand{\FigureVLCA}{
\begin{tikzpicture}[baseline={([yshift=-.8ex]current bounding box.center)},scale=1.2]
    \draw [white] (0.5, -0.2) -- (1.5, -0.2); 
    \fill [red, opacity = 0.2] (0.0, 0.0) -- (0.0, 1.6) -- (0.9, 1.6) -- (0.9, 0.0);
    \fill [black, opacity = 0.2] (0.9, 0.0) -- (0.9, 1.6) -- (1.1, 1.6) -- (1.1, 0.0);
    \fill [blue, opacity = 0.2] (1.1, 0.0) -- (1.1, 1.6) -- (2.0, 1.6) -- (2.0, 0.0);
    \draw [ultra thick] (1.0, 0.8) circle (0.6);
    \draw [ultra thick] (1.0, 0.8) circle (0.4);
    \node [centered, rotate=0.] at (0.25,0.8) {\scriptsize $a$};
    \node [centered, rotate=0.] at (0.75,0.8) {\scriptsize $b$};
    \node [centered, rotate=0.] at (1.25,0.8) {\scriptsize $s$};
    \node [centered, rotate=0.] at (1.75,0.8) {\scriptsize $r$};
\end{tikzpicture}
}

\newcommand{\FigureVLCB}{
\begin{tikzpicture}[baseline={([yshift=-.8ex]current bounding box.center)},scale=1.2]
    \draw [white] (0.5, -0.2) -- (1.5, -0.2); 
    \fill [red, opacity = 0.2] (0.0, 0.0) -- (0.0, 1.6) -- (0.9, 1.6) -- (0.9, 0.0);
    \fill [black, opacity = 0.2] (0.9, 0.0) -- (0.9, 1.6) -- (1.1, 1.6) -- (1.1, 0.0);
    \fill [blue, opacity = 0.2] (1.1, 0.0) -- (1.1, 1.6) -- (2.0, 1.6) -- (2.0, 0.0);
    \draw [ultra thick] (1.0, 1.2) circle (0.25);
    \draw [ultra thick] (1.0, 0.4) circle (0.25);
    \draw [ultra thick] (0.76,0.367) arc [start angle=240, end angle=120, x radius=0.5, y radius=0.5];
    \draw [ultra thick] (1.24,0.367) arc [start angle=-60, end angle=60, x radius=0.5, y radius=0.5];
    \node [centered, rotate=0.] at (0.35,0.8) {\scriptsize $c$};
    \node [centered, rotate=0.] at (1.65,0.8) {\scriptsize $t$};
    \node [centered, rotate=0.] at (0.75,1.5) {\scriptsize $a$};
    \node [centered, rotate=0.] at (0.75,0.9) {\scriptsize $b$};
    \node [centered, rotate=0.] at (0.9,1.2) {\scriptsize $\nu$};
    \node [centered, rotate=0.] at (0.75,0.7) {\scriptsize $b$};
    \node [centered, rotate=0.] at (0.75,0.1) {\scriptsize $a$};
    \node [centered, rotate=0.] at (0.9,0.4) {\scriptsize $\nu$};
    \node [centered, rotate=0.] at (1.25,1.5) {\scriptsize $r$};
    \node [centered, rotate=0.] at (1.25,0.9) {\scriptsize $s$};
    \node [centered, rotate=0.] at (1.1,1.2) {\scriptsize $\mu$};
    \node [centered, rotate=0.] at (1.25,0.7) {\scriptsize $s$};
    \node [centered, rotate=0.] at (1.25,0.1) {\scriptsize $r$};
    \node [centered, rotate=0.] at (1.1,0.4) {\scriptsize $\mu$};
\end{tikzpicture}
}

\newcommand{\FigureVLCC}{
\begin{tikzpicture}[baseline={([yshift=-.8ex]current bounding box.center)},scale=1.2]
    \draw [white] (0.5, -0.2) -- (1.5, -0.2); 
    \fill [red, opacity = 0.2] (0.0, 0.0) -- (0.0, 1.6) -- (0.9, 1.6) -- (0.9, 0.0);
    \fill [black, opacity = 0.2] (0.9, 0.0) -- (0.9, 1.6) -- (1.1, 1.6) -- (1.1, 0.0);
    \fill [blue, opacity = 0.2] (1.1, 0.0) -- (1.1, 1.6) -- (2.0, 1.6) -- (2.0, 0.0);
    \draw [ultra thick] (1.0, 0.8) circle (0.6);
    \node [centered, rotate=0.] at (0.25,0.8) {\scriptsize $c$};
    \node [centered, rotate=0.] at (1.75,0.8) {\scriptsize $t$};
\end{tikzpicture}
}